\documentclass[aps,prb,superscriptaddress,amsfonts,amsmath,amssymb,floatfix,showpacs,reprint,longbibliography,footinbib]{revtex4-2}%

\usepackage{graphicx}
\usepackage{svg}
\usepackage{amsmath,amssymb}
\usepackage{dcolumn}
\usepackage{bm}
\usepackage{color}
\usepackage{hyperref}
\hypersetup{colorlinks=true,breaklinks,linkcolor=blue,urlcolor=blue,citecolor=blue}
\usepackage{xspace}
\usepackage{multirow}
\usepackage{physics}
\usepackage[version=4]{mhchem}
\usepackage{booktabs}
\usepackage[normalem]{ulem}

\newcommand{\cerhas}{\ce{CeRh2As2}}
\newcommand{\Gsevenone}{\ensuremath{\Gamma_7^{(1)}}}
\newcommand{\Gseventwo}{\ensuremath{\Gamma_7^{(2)}}}
\newcommand{\Gseven}{\ensuremath{\Gamma_7}}
\newcommand{\Gsix}{\ensuremath{\Gamma_6}}

\newcommand{\QA}{\bm{Q}_\mathrm{A}}
\newcommand{\QM}{\bm{Q}_\mathrm{M}}
\newcommand{\QZ}{\bm{Q}_\mathrm{Z}}
\newcommand{\QG}{\bm{Q}_\mathrm{\Gamma}}

\begin{document}

\title{Multipolar fluctuations from localized $4f$ electrons in {\cerhas}}

\author{Koki Numa}
\affiliation{Department of Physics, Okayama University, Okayama 700-8530, Japan}
\author{Eri Matsuda}
\affiliation{Department of Physics, Okayama University, Okayama 700-8530, Japan}
\author{Akimitsu Kirikoshi}
\affiliation{Research Institute for Interdisciplinary Science, Okayama University, Okayama 700-8530, Japan}
\author{Junya Otsuki}
\affiliation{Research Institute for Interdisciplinary Science, Okayama University, Okayama 700-8530, Japan}

\date{\today}

\begin{abstract}
The heavy-fermion superconductor {\cerhas} exhibits a non-superconducting phase transition that precedes the emergence of superconductivity.
The nature of the corresponding order parameter remains under debate, with competing proposals involving magnetic dipoles or electric quadrupoles.
We derive the momentum-dependent multipolar susceptibilities and effective interactions among the localized $4f$ electrons, based on the framework of density functional theory combined with dynamical mean-field theory.
Magnetic fluctuations within the crystalline-electric-field (CEF) ground-state doublet are dominated by $\bm{q}=(1/2,1/2,0)$, corresponding to a two-dimensional checkerboard configuration of the magnetic moment $M_z$ along the $c$ axis.
Hybridization between the CEF ground state and the first-excited doublet gives rise to leading magnetic octupole fluctuations of $z(x^2-y^2)$ symmetry, followed by electric quadrupole fluctuations of $x^2-y^2$ and $\{ yz, zx \}$ symmetries.
By taking into account the anisotropic magnetic-field dependence of the transition temperature $T_0$, we conclude that an antiferromagnetic order of $M_z$ at $\bm{q}=(1/2,1/2,0)$ is consistent with the experiments, owing to the enhancement of $T_0$ caused by fluctuations of the field-induced quadrupole of $\{ yz, zx \}$ type under an in-plane magnetic field.
\end{abstract}

\maketitle

\section{Introduction}
\label{sec:introduction}

Cerium-based materials exhibit superconductivity, magnetism, and multipolar ordering.
Their distinctive properties often originate from the large total angular momentum $j=5/2$ arising from the strong spin-orbit coupling.
The absence of local inversion symmetry can further give rise to exotic electronic states such as odd-parity order parameters.
{\cerhas} is one such material that has recently attracted considerable attention.

\begin{figure}[b]
    \includegraphics[width=0.9\linewidth]{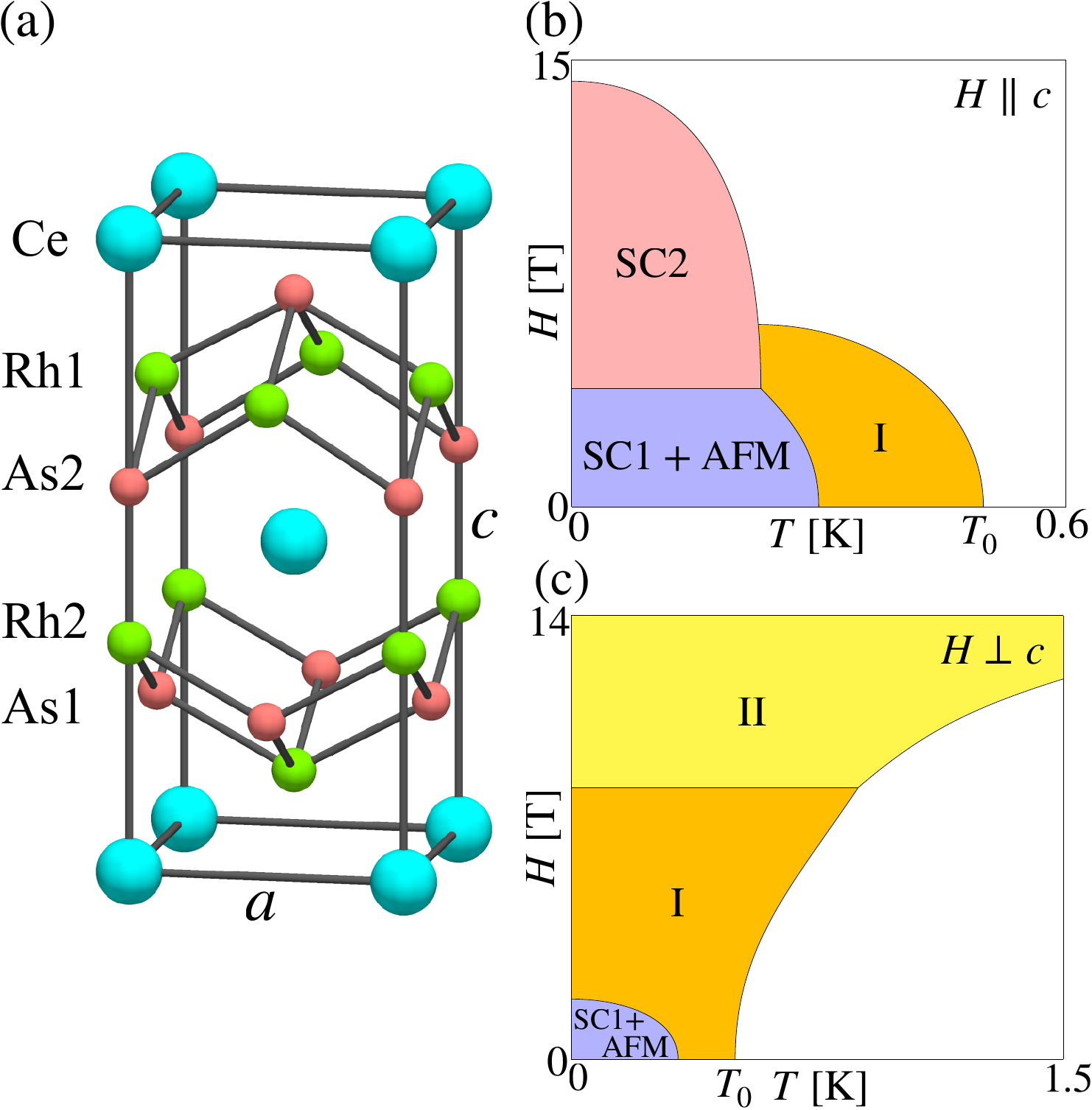}
    \caption{(a) Crystal structure of {\cerhas} and schematic $T$--$H$ phase diagrams based on experimental results in the magnetic field applied along (b) the $c$ axis~\cite{Semeniuk2023} and (c) the in-plane direction~\cite{Hafner2022}.}
    \label{fig:str_phase}
\end{figure}

{\cerhas} is a heavy-fermion superconductor with a critical temperature $T_{\mathrm{SC}}\approx0.3~\mathrm{K}$ and a large Sommerfeld coefficient of $700$--$1200~\mathrm{mJ/mol\,K^{2}}$~\cite{Khim2021,Semeniuk2023}.
Its crystal structure is of $\mathrm{CaBe_2Ge_2}$-type, belonging to the nonsymmorphic space group $P4/nmm$ ($D^7_{4h}$, No.~129).
This structure is closely related to the well-known $\mathrm{ThCr_2Si_2}$-type structure in space group $I4/mmm$ ($D^{17}_{4h}$, No.~139), but lacks local inversion symmetry at the Ce site as illustrated in Fig.~\ref{fig:str_phase}(a).
A notable feature of this compound is that the superconducting state persists up to $H_{c2}\approx 14~\mathrm{T}$ for a magnetic field along the $c$ axis, which is anomalously high relative to the energy scale of $T_{\mathrm{SC}}\approx 0.3~\mathrm{K}$~\cite{Khim2021}.
The superconducting phase is divided by a first-order phase transition at $H\approx 3.9~\mathrm{T}$ into a low-field phase (SC1) and a high-field phase (SC2)~\cite{Khim2021, Semeniuk2023, Hafner2022, Landaeta2022-angle}, as schematically illustrated in Fig.~\ref{fig:str_phase}(b).
Moreover, antiferromagnetism (AFM) coexists with the superconductivity in SC1, forming SC1+AFM phase, as suggested by nuclear quadrupole resonance (NQR)~\cite{Kibune2022} and muon spin relaxation ($\mu$SR)~\cite{Khim2025-musr}.
The superconducting properties of this material have been extensively investigated through the nuclear magnetic resonance (NMR) and NQR~\cite{Kitagawa2022,Kibune2022,Ogata2023,Ogata2024,Ogata2026-larh2as2}, optical conductivity~\cite{Kimura2021}, thermal conductivity~\cite{Onishi2022}, angle-resolved photoemission spectroscopy (ARPES)~\cite{Chen2024-ARPES+U,Chen2024-ARPES}, inelastic neutron scattering~\cite{Chen2024-neutron}, an external pressure~\cite{Siddiquee2023,Semeniuk2024}, point contact spectroscopy~\cite{dong_pairing_strength}, and electronic structure calculations~\cite{Nogaki2021,Ali2023,Ishizuka2024,Wu2024,ma_DFT+DMFT_copmression}.
Theoretical investigations address superconductivity influenced by, for example, the nonsymmorphic structure and the Rashba spin-orbit coupling caused by non-centrosymmetry of correlated sites~\cite{Yoshida2012,Mockli2021,Schertenleib2021,Skurativska2021,Ptok2021,Machida2022,Cavanagh2022,Mockli2022,Hazra2023,cavanagh2023,Suh2023,nogaki_multi_fluc,Szabo2024,amin_kramers_magsc,minamide_meron,lee_unified_picture}.

Another intriguing property in {\cerhas} is a non-superconducting phase transition at $T_0\approx 0.5~\mathrm{K}$, detected in specific heat, thermal expansion, and resistivity measurements~\cite{Khim2021,Semeniuk2023,Mishra2022,Pfeiffer2024}.
The phase in the range $T_{\rm{SC}} < T < T_0$ is referred to as phase~I.
The absence of anomalies at $T_0$ in the AC magnetic susceptibility and in NQR and NMR spectra suggests a nonmagnetic origin of the transition~\cite{Kibune2022, Ogata2024}, although recent experiments for a high-quality sample indicate the emergence of antiferromagnetic moments in phase~I~\cite{Chajewski2024,Khim2025-musr,khanenko_outmag}.
A striking aspect of this transition is its strongly anisotropic response to the magnetic field:
$T_0$ is strongly suppressed by $H \parallel c$~\cite{Khim2021,Semeniuk2023,khanenko_outmag}, while $T_0$ is enhanced by the in-plane magnetic field~\cite{Chajewski2024, Hafner2022} as shown in Figs.~\ref{fig:str_phase}(b) and \ref{fig:str_phase}(c).
The magnetic-field enhancement of the transition temperature is reminiscent of \ce{CeB6}, in which the antiferro-quadrupolar order is stabilized by the field-induced magnetic octupoles, although no such anisotropy is present there~\cite{Shiina1997,Kuramoto-review}.
Resistivity measurements in {\cerhas} further suggest a transition into another phase, termed phase~II, under magnetic field $H \perp c$~\cite{Hafner2022}.
Furthermore, a non-Fermi-liquid behavior has been observed in the paramagnetic state above $T=T_0$~\cite{Khanenko2025,Miyake2024}.

Two scenarios have been considered as the origin of phase~I.
Hafner \textit{et al.} proposed a quadrupole density wave (QDW) order that may emerge due to a nesting of the Fermi surfaces obtained from the renormalized band structure calculations~\cite{Hafner2022, Zwicknagl1992}.
The crystalline-electric-field (CEF) splitting divides the sextet of $j=5/2$ into three doublets, with the ground-state and first-excited doublets separated by $\Delta_1=30~\mathrm{K}$.
Quadrupole degrees of freedom become active only when the first-excited doublet is involved, which can occur under sufficiently strong hybridization.
Indeed, the Kondo temperature is estimated to be $T_{\mathrm{K}}\approx 30$~K~\cite{Khim2021}, comparable to $\Delta_1$.
The x-ray absorption spectra indicate the formation of a quasiquartet~\cite{Cristovam2024}.
Symmetry analyses have also discussed possible quadrupolar order parameters~\cite{Harima2023}.

On the other hand, Schmidt and Thalmeier discussed an antiferromagnetic order~\cite{Schmidt2024,Thalmeier_thermodynamic}.
In their scenario, the observed anisotropy with respect to the applied magnetic field is attributed to fluctuations of quadrupole moments induced by the magnetic field.
By introducing interactions between $xy$-type quadrupoles, they demonstrated that the anisotropy of the experimental phase diagram can be reproduced within an antiferromagnetic ordering of in-plane moments.
We note, however, that the antiferromagnetic and quadrupolar interactions are introduced \textit{ad hoc}.

In this paper, we address the order parameter in phase~I of {\cerhas}, taking
the electronic structure and strong correlations among $4f$ electrons into account.
Our strategy consists of two steps.
First, we derive the multipolar susceptibilities and effective interactions using density functional theory combined with dynamical mean-field theory (DFT+DMFT).
A few candidate order parameters are listed at this stage.
Next, we examine the anisotropy of the transition temperature. Combining a phenomenological argument with the DFT+DMFT results, we identify an order parameter that simultaneously exhibits strong fluctuations and reproduces the observed anisotropy.
In this way, we propose a two-dimensional antiferromagnetic order of magnetic moments parallel to the $c$ axis with $\bm{q}=(1/2,1/2,0)$ for the primary candidate, and an antiferromagnetic order of in-plane moments with $\bm{q}=(0,0,1/2)$ as a subleading candidate.

This paper is organized as follows.
Section~\ref{sec:Results} presents a construction of a low-energy model based on DFT calculations and discusses the single-particle excitation spectrum by the DFT+DMFT method.
In Sec.~\ref{sec:multi-chi}, we derive the momentum-dependent multipolar susceptibilities and effective interactions.
Taking the anisotropy of the transition temperature into account, we identify the order parameter in Sec.~\ref{sec:anisotropy}.
Relations to experimental results and previous theoretical proposals are discussed in Sec.~\ref{sec:Discussion}.
The paper is summarized in Sec.~\ref{sec:Summary}.

\section{Electronic structure}
\label{sec:Results}

\subsection{DFT calculations}
\label{DFT-cal}

\begin{figure}[t]
\centering
    \includegraphics[width=1\linewidth]{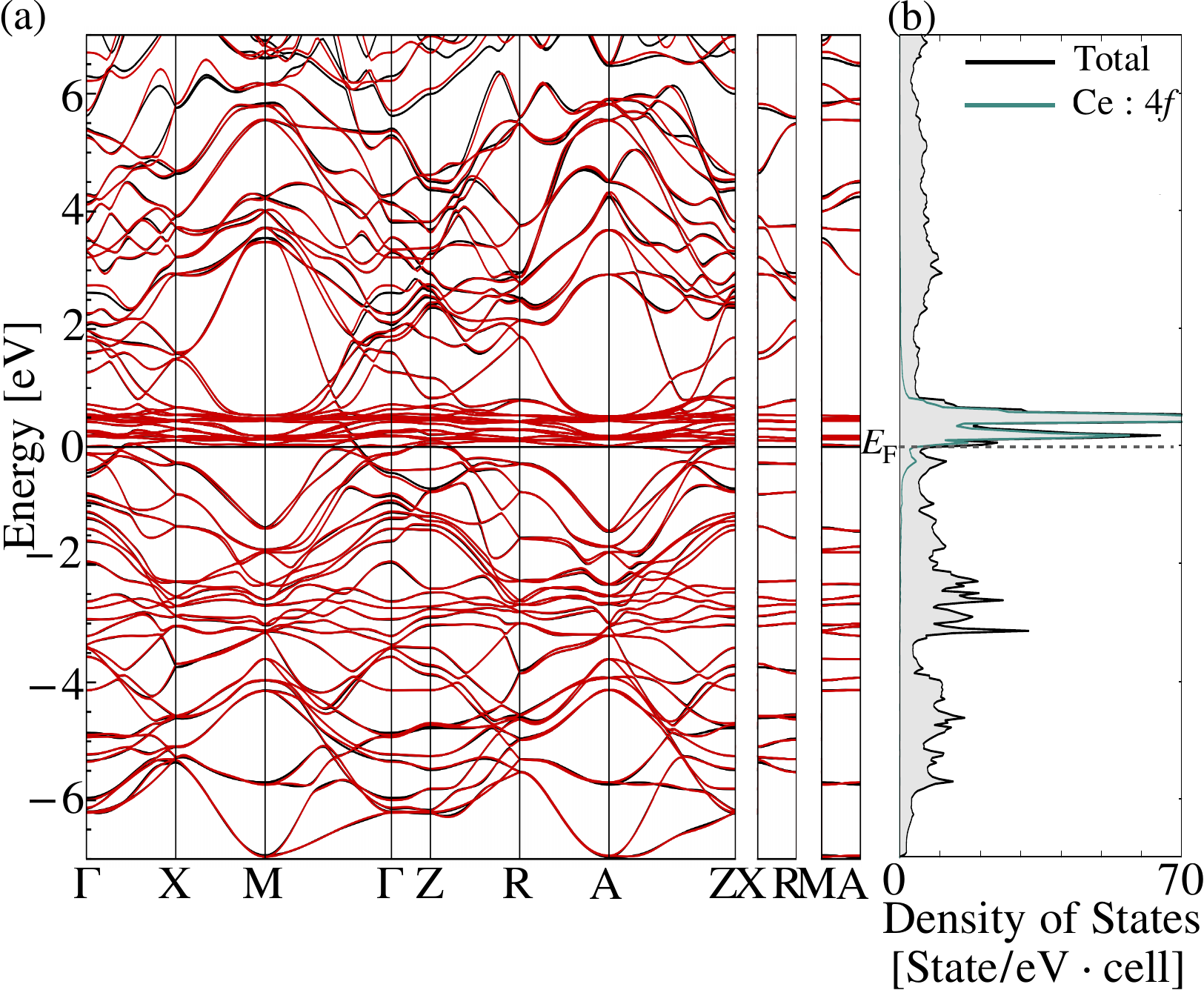}
    \includegraphics[width=0.55\linewidth]{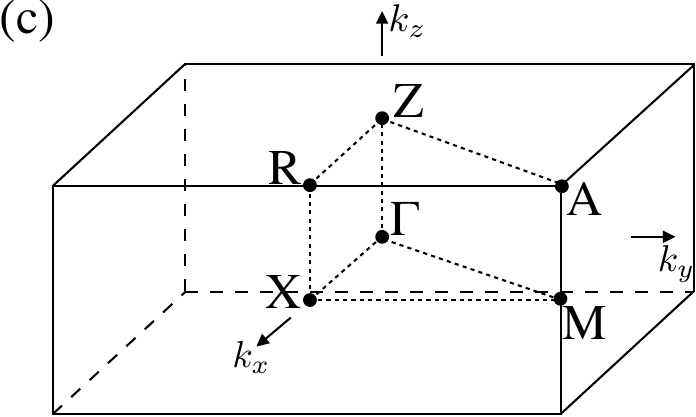}
    \caption{(a) Fully relativistic band structure of {\cerhas} calculated using the FPLO package. The black lines represent the DFT result, and the red lines represent the tight-binding bands obtained from projective Wannier functions. (b) Density of states. The black and green lines show the total and Ce $4f$ contributions, respectively. (c) The high-symmetry points in the Brillouin zone.}
    \label{fig:dft-band}
\end{figure}

We begin by calculating the electronic structure of {\cerhas}.
The structure parameters are provided in Ref.~\cite{Khim2021}.
We employ the full-potential local orbitals (FPLO) package \cite{Koepernik1999-FPLO,Opahle1999-FPLO} with the generalized gradient approximation to calculate a fully relativistic band structure.
The Brillouin zone is discretized into $48\times48\times24$ $\bm{k}$ points.
$4f$ electrons of Ce ions are treated as itinerant.
Figure~\ref{fig:dft-band} shows the DFT band structure (black lines) and density of states.
This calculated electronic structure is in close agreement with the previous full-potential linearized augmented-plane-wave (FLAPW) results for \cerhas~\cite{Nogaki2021}.
To construct a low-energy model, we employ symmetry-protected maximally projected Wannier functions \cite{Eschrig2009-wannier, Koepernik2023-wannier}, yielding a 124-band tight-binding model consisting of Ce $4f,5d$ and $6s$, Rh $4d$ and $5s$, and As $4p$ orbitals.
The resulting tight-binding bands [red lines in Fig.~\ref{fig:dft-band}(a)] reproduce the DFT dispersion accurately.

\begin{figure}[t]
\centering
    \includegraphics[width=0.7\linewidth]{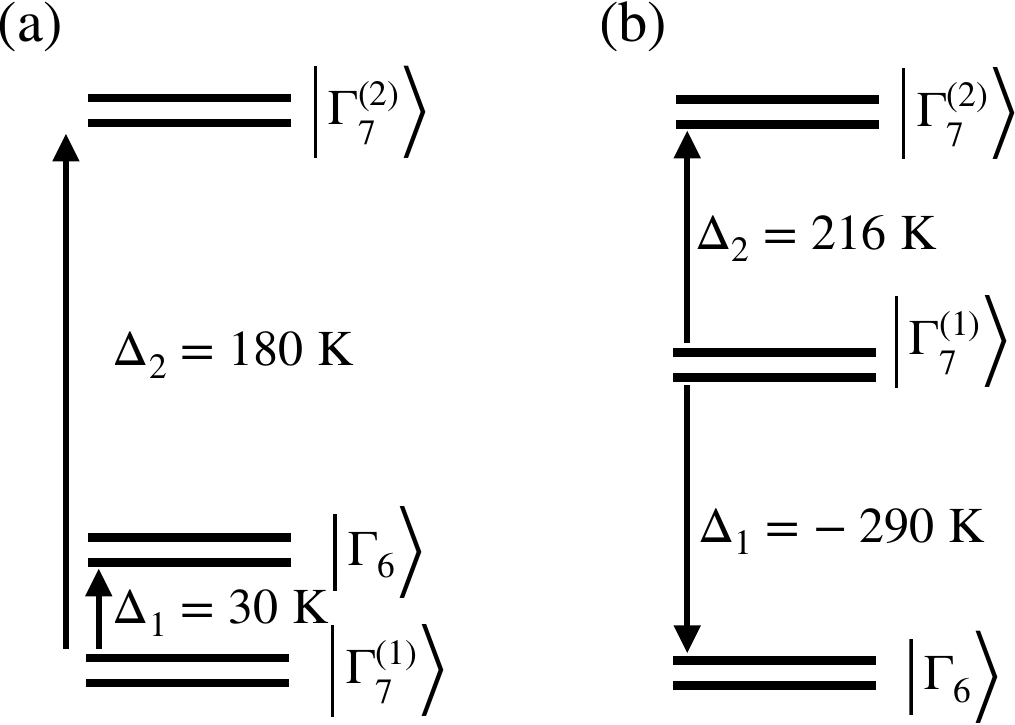}
    \caption{The CEF level schemes of $j=5/2$ of $4f$ electrons. (a)~The experimental one~\cite{Hafner2022} and (b) the DFT result. $\Delta_1$ and $\Delta_2$ denote the energy of the {\Gsix} and {\Gseventwo} states, respectively, relative to the {\Gsevenone} state.}
    \label{fig:cef_split}
\end{figure}

We discuss the crystalline-electric field (CEF) states of $4f$ electrons.
The $j=5/2$ sextet splits into three Kramers doublets under the CEF with point-group symmetry $C_{4v}$.
From the magnetic susceptibility and specific-heat measurements, the CEF level scheme shown in Fig.~\ref{fig:cef_split}(a) has been proposed~\cite{Hafner2022}.
The ground state is the {\Gseven} doublet, which is denoted by {\Gsevenone}.
The first-excited state is {\Gsix} doublet, separated by $\Delta_1=30$~K.
The second excited state is another {\Gseven}, labeled {\Gseventwo}, lying $\Delta_2=180$~K above the ground state.

Our DFT results yield the CEF level scheme shown in Fig.~\ref{fig:cef_split}(b).
The two low-lying doublets are inverted compared to the experimental scheme.
The energy splitting obtained is $\Delta_1=-290$~K (the negative sign indicating the reversed order of {\Gsevenone} and {\Gsix}).
Since the CEF states play a dominant role in subsequent calculations of multipolar fluctuations, we artificially adjust the CEF potential to reproduce the experimental CEF level scheme.

\subsection{DMFT calculations}
\label{DMFT-cal}

We treat the strong correlations among $4f$ electrons in the DMFT~\cite{Georges1996,Kotliar2006}.
For simplicity, we omit the $j=7/2$ states of $4f$ electrons, which lie $\Delta_{\mathrm{SOC}} \simeq 0.33$~eV above the $j=5/2$ states due to the spin-orbit coupling.
The single-particle Green's function for the remaining 108 orbitals is given by
\begin{align}
    \hat{G}(\bm{k},i\omega_n)
    = \Big[ &(i\omega_n+\mu)\hat{I}-\hat{H}_{\mathrm{DFT}}(\bm{k})-\hat{H}_{\mathrm{CEF}}
    \notag \\
    &-\hat{\Sigma}_{\mathrm{loc}}(i\omega_n)-\varepsilon_f\hat{P}_f \Big]^{-1},
\end{align}
where the hat denotes a matrix having site, spin, and orbital indices within a unit cell.
$\hat{I}$ is the identity matrix.
$\hat{H}_{\mathrm{DFT}}(\bm{k})$ is the tight-binding Hamiltonian obtained by the DFT calculations, where the CEF potential for $4f$ electrons is removed.
$\hat{H}_{\mathrm{CEF}}$ is the CEF potential that reproduces the experimental CEF level scheme in Fig.~\ref{fig:cef_split}(a).
We adopt this simple approach, although enforcing charge self-consistency between DFT and DMFT would improve the CEF level scheme~\cite{Delange2017,Pourovskii2020}, which could allow us to avoid the manual adjustment of the CEF potential.
$\hat{\Sigma}_{\mathrm{loc}}(i\omega_n)$ is the local self-energy in the DMFT.
$\varepsilon_f$ is the energy of the $4f$ levels. $\hat{P_f}$ denotes a projection operator onto $4f$ orbitals. This term works as a double-counting correction between DFT and DMFT calculations.

We compute the self-energy $\hat{\Sigma}_{\mathrm{loc}}$ using the exact diagonalization within the Hubbard-I approximation.
Here, we adopt the fully rotationally invariant Coulomb interaction with the conventional parametrization~\cite{Anisimov1997}.
The interaction parameters together with $\varepsilon_f$ are chosen to reproduce the experimental spectrum as follows.
The $4f$ level is set to $\varepsilon_f=-1.65$~eV to reproduce the ARPES experiments, which show the $4f^1\rightarrow 4f^0$ excitation at $\Delta_-=1.5~\mathrm{eV}$ below the Fermi level~\cite{Chen2024-ARPES}.
The direct Coulomb interaction is taken as $U=5.8$\,eV with the Hund's coupling $J=0.8$~eV~\cite{Locht2016}, to reproduce the $4f^1\rightarrow 4f^2$ excitation energy $\Delta_+=3.1$~eV for elemental Ce~\cite{Herbst1978}.

\begin{figure}[t]
    \includegraphics[width=1\linewidth]{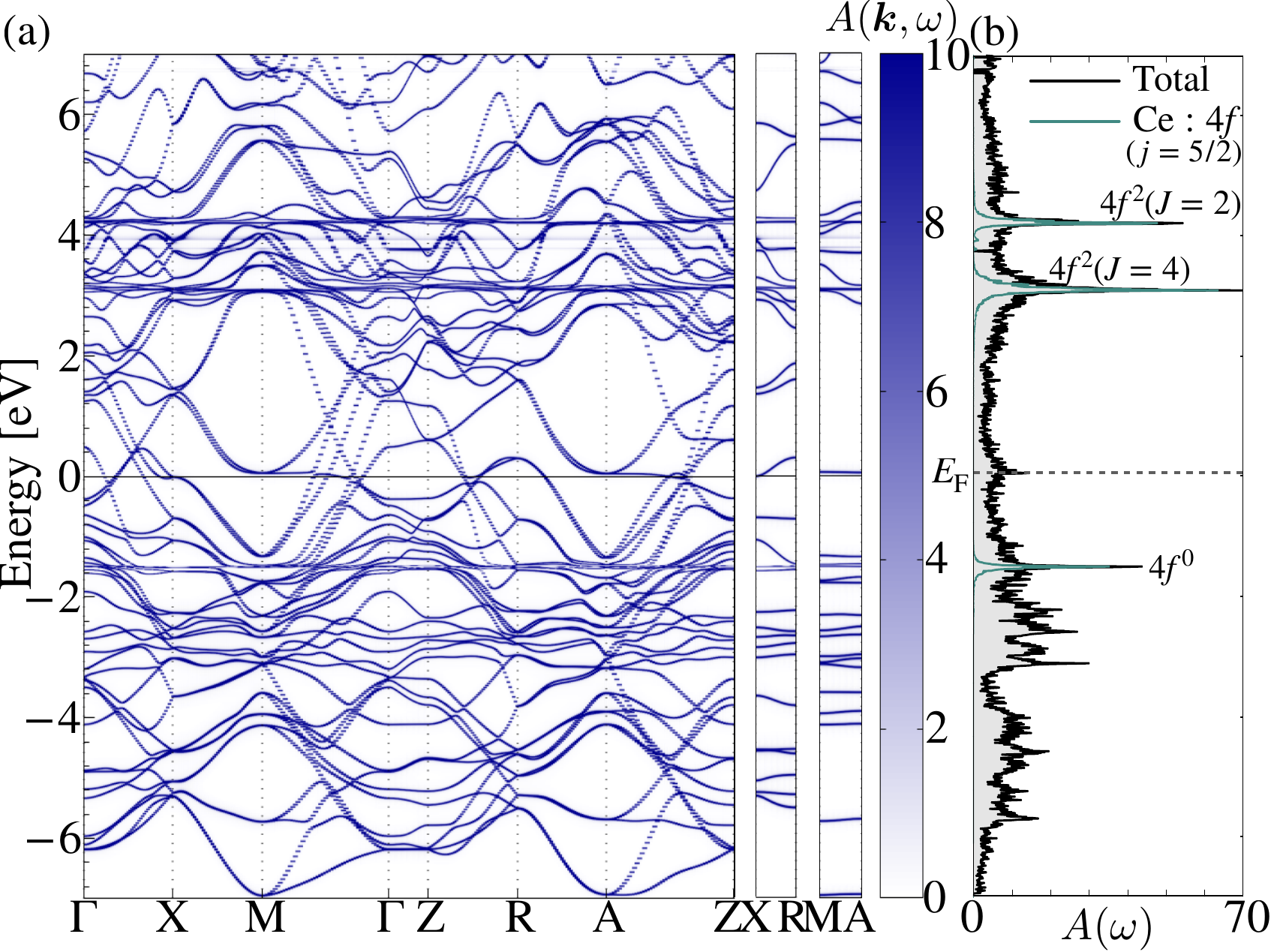}
    \caption{(a) The single-particle excitation spectrum $A(\bm{k},\omega)$ and (b) the $\bm{k}$-integrated spectrum $A(\omega)$ computed using the DFT+DMFT method.}
    \label{fig:akw}
\end{figure}

Figure~\ref{fig:akw} shows the single-particle excitation spectrum thus obtained in DFT+DMFT calculations.
The itinerant $4f$ band near $E_{\mathrm{F}}$ in the DFT [Fig.~\ref{fig:dft-band}(a)] is now located around $-1.5~\mathrm{eV}$ and $3.1$\,eV, indicating a localized nature of $4f$ electrons.
The resultant band structure near $E_{\mathrm{F}}$ agrees with the DFT results for $\mathrm{LaRh_2As_2}$~\cite{Landaeta2022}.
In particular, our results well reproduce the van Hove singularity near the Fermi level at the X point observed in the ARPES measurements~\cite{Chen2024-ARPES+U}.

\section{Multipolar susceptibilities and interactions}
\label{sec:multi-chi}

\subsection{Method}
We consider spin, charge, and orbital fluctuations within the DMFT.
The momentum-dependent susceptibility is defined by
\begin{align}
    \chi_{i m_1m_2, jm_3m_4}(\bm{q})=\int^{\beta}_{0}d\tau \left< O_{i m_1m_2}(\bm{q},\tau)O_{j m_3m_4}(-\bm{q})\right>,
\end{align}
where $O_{i m_1m_2}(\bm{q})$ is the Fourier transform of the local density operator
\begin{align}
    O_{i m_1m_2}(\bm{R})=f_{i m_1}^{\dag}(\bm{R}) f_{i m_2}(\bm{R}).
\end{align}
Here, $\bm{R}$ are the lattice vectors locating the origin of each unit cell, $i$ labels the site in the unit cell, and $m_1$ denotes the $z$ component of $j=5/2$.

In DMFT, $\chi(\bm{q})$ can be computed by solving the Bethe-Salpeter equation for the two-particle Green's function~\cite{Georges1996}. In this paper, we employ an alternative approximate formula that avoids the explicit calculation of the two-particle vertex~\cite{Otsuki2019-SCL}.
In this strong-coupling limit (SCL) formula, $\chi(\bm{q})$ is given by
\begin{align}
    \hat{\chi}(\bm{q}) = [\hat{\chi}_\mathrm{loc}^{-1} - \hat{I}(\bm{q})]^{-1},
    \label{eq:chi_q_matrix}
\end{align}
where the hat denotes a matrix with respect to the combined indices $i m_1 m_2$ and $j m_3 m_4$.
Here, $\hat{\chi}_\mathrm{loc}$ is the local susceptibility evaluated from the effective impurity problem in DMFT, and $\hat{I}(\bm{q})$ describes the effective intersite interactions.
The latter is computed from $\hat{G}(\bm{k},i\omega_n)$ together with a function $\phi(i\omega_n)$, which effectively replaces the vertex function.
We adopt the two-pole approximation for $\phi(i\omega_n)$ (referred to as the SCL3 scheme), in which $\phi(i\omega_n)$ is represented in terms of two poles located at $\Delta_-$ and $\Delta_+$, corresponding to the excitation energies from $4f^1$ configuration to $4f^0$ and $4f^2$, respectively~\cite{Otsuki2019-SCL}.
This scheme successfully reproduces the antiferro-quadrupolar order in \ce{CeB6} and the ferromagnetism in \ce{CeRh6Ge4}~\cite{Otsuki2024-ceb6,cerh6ge4-itokazu}.
A related approach based on the Hubbard-I approximation has also been developed~\cite{Pourovskii2016,Pourovskii2019}.

Here, we comment on the influence of the $j=7/2$ states eliminated in the DMFT calculations. The inclusion of $j=7/2$ is necessary to reproduce the $4f^2$ multiplet structure more accurately, which appears above 3\,eV in $A(\bm{k},\omega)$ in Fig.~\ref{fig:akw}. Within the SCL formula with the two-pole approximation, the $4f^2$ configuration is represented by a single pole located at $\omega=\Delta_+$, thereby neglecting the splitting of the $4f^2$ configuration. Consequently, incorporating the $j=7/2$ states does not affect the results in the present scheme.

After all components of $\chi_{i m_1m_2, jm_3m_4}(\bm{q})$ are obtained, we diagonalize it as follows:
\begin{align}
   \chi^{(\xi)}(\bm{q})
   &=\sum^{}_{ij}\sum^{}_{m_1,\cdots,m_4}
   \notag \\
   &\times u^{(\xi)}_{i m_1m_2}(\bm{q}) \chi_{i m_1m_2, j m_3m_4}(\bm{q}) u^{(\xi)}_{j m_3m_4}(\bm{q})^{\ast}.
\end{align}
Here, eigenmodes are labeled with the superscript $\xi$.
The eigenvalues $\chi^{(\xi)}(\bm{q})$ represent the magnitude of the fluctuations and the eigenvectors $u^{(\xi)}_{i m_1m_2}(\bm{q})$ represent the corresponding magnetic and orbital configuration.

The character of the fluctuations is identified as follows.
We first construct the symmetry-adapted multipole basis (SAMB) $O^{(\gamma)}$ using open-source software MultiPie
~\cite{multipie} as
\begin{align}
    O^{(\gamma)} = \sum_{i mm'} {z^{(\gamma)}_{imm'}} O_{imm'}.
\end{align}
Each basis $\gamma$ belongs to one of the irreducible representations in the crystallographic point group.
We then project the eigenvector
$u^{(\xi)}_{i m_1m_2}(\bm{q})$ onto the SAMB by
\begin{align}
    c^{(\xi)}_{\gamma}(\bm{q}) = \sum_{i m_1 m_2} {z^{(\gamma) \ast}_{i m_1 m_2}} u^{(\xi)}_{i m_1m_2}(\bm{q}).
\end{align}
At high-symmetry $\bm{q}$ points defined in Fig.~\ref{fig:dft-band}(c), the coefficients $c^{(\xi)}_{\gamma}(\bm{q})$ become finite exclusively in one irreducible representation.
In contrast, at generic $\bm{q}$ points, the coefficients belonging to different irreducible representations can take finite values.
We therefore label each eigenmode $\xi$ by the irreducible representation $\gamma$ that has the largest weight $|c^{(\xi)}_{\gamma}(\bm{q})|^2$.

\subsection{Multipoles}

We perform a symmetry analysis of possible fluctuation modes using multipoles, which enables systematic characterization of symmetry-broken states~\cite{Kuramoto-review} and the associated cross-correlated responses~\cite{Hayami-review}.
In the group theory, the operators within the {\Gseven} doublet are decomposed as follows:
\begin{align}
    \Gamma_7 \otimes \Gamma_7 = A_{1}^{+} \oplus A_{2}^{-} \oplus E^{-},
\end{align}
where the superscript $+$ ($-$) indicates the time-reversal-even (odd) component, which appears by a symmetric (antisymmetric) product between two irreducible representations~\cite{inui2012group}.
These operators are summarized in Table~\ref{tab:atomic_multipoles} (the second column).
$A_1^+$ corresponds to the charge operator $Q_0$, and $A_2^-$ and $E^-$ correspond to magnetic dipoles $M_z$ and $\{ M_x, M_y \}$, respectively.

\begin{table}[t]
    \caption{Atomic multipoles of $4f$ electrons. The column ``Irrep'' indicates the irreducible representation of the $C_{4v}$ point group. The second and third columns show the four multipoles in the {\Gseven} doublet system and the additional 12 multipoles that appear in the {\Gseven}--{\Gsix} pseudo-quartet system, respectively. The notation follows Ref.~\cite{Hayami-review}.}
    \label{tab:atomic_multipoles}
    \begin{tabular}{llll}\hline
        Irrep & $\Gamma_7 \otimes \Gamma_7$ & $\Gamma_7 \otimes \Gamma_6$ \\
        \hline
        $A_{1}^{+}$ & $Q_{0}$ & $Q_{3z^2-r^2}$ \\
        $A_{2}^{+}$ & &  \\
        $B_{1}^{+}$ & & $Q_{x^2-y^2}$ \\
        $B_{2}^{+}$ & & $Q_{xy}$ \\
        $E^{+}$ & & $\{ Q_{yz},\ Q_{zx} \}$ \\
        \hline
        $A_{1}^{-}$ & &  \\
        $A_{2}^{-}$ & $M_{z}$ & $M_{z(5z^2-3r^2)}$ \\
        $B_{1}^{-}$ & & $M_{xyz}$ \\
        $B_{2}^{-}$ & & $M_{z(x^2-y^2)}$ \\
        $E^{-}$ & $\{ M_{x},\ M_{y} \}$ & $\{M_{x(x^2-z^2)},\ M_{y(y^2-z^2)} \}$,\\
        && $\{ M_{x(5x^2-3r^2)},\ M_{y(5y^2-3r^2)} \}$ \\
        \hline
    \end{tabular}
\end{table}

We next consider multipoles that appear by mixing of the ground-state doublet {\Gsevenone} and the first-excited doublet {\Gsix}.
The symmetries of the multipole operators are represented as
\begin{align}
    \Gamma_7 \otimes \Gamma_6 = B_{1}^{\pm} \oplus B_{2}^{\pm} \oplus E^{\pm},
\end{align}
where the superscript $\pm$ indicates that both electric and magnetic operators exist.
The third column of Table~\ref{tab:atomic_multipoles} summarizes these operators.
The operators in $B_{1}^{+}$, $B_{2}^{+}$, and $E^{+}$ are electric quadrupoles of $Q_{x^2-y^2}$, $Q_{xy}$, and $\{ Q_{yz}, Q_{zx} \}$, respectively.
The remaining quadrupole, namely $Q_{3z^2-r^2}$, appears as a higher-order charge distribution in $A_{1}^+$ representation.
The magnetic operators in $B_{1}^{-}$, $B_{2}^{-}$, and $E^{-}$ correspond to magnetic octupoles.
In addition, octupole operators also appear as higher-rank operators of $M_x$, $M_y$, and $M_z$.

Thus, within the {\Gsevenone} doublet, there are four multipoles (charge and magnetic dipoles), while the {\Gsevenone}--{\Gsix} pseudo-quartet system additionally hosts five electric quadrupoles and seven magnetic octupoles.
The two Ce sites in a unit cell allow either ferroic (F) or antiferroic (AF) configurations.
Hereafter, we represent the combined atomic multipoles and the site degrees of freedom, for example, by $M_z$(F) and $M_z$(AF).
For reference, we provide a symmetry classification based on cluster multipoles in Appendix~\ref{app:multipole}.

\subsection{Results for susceptibilities}

\begin{figure}[t]
    \includegraphics[width=1\linewidth]{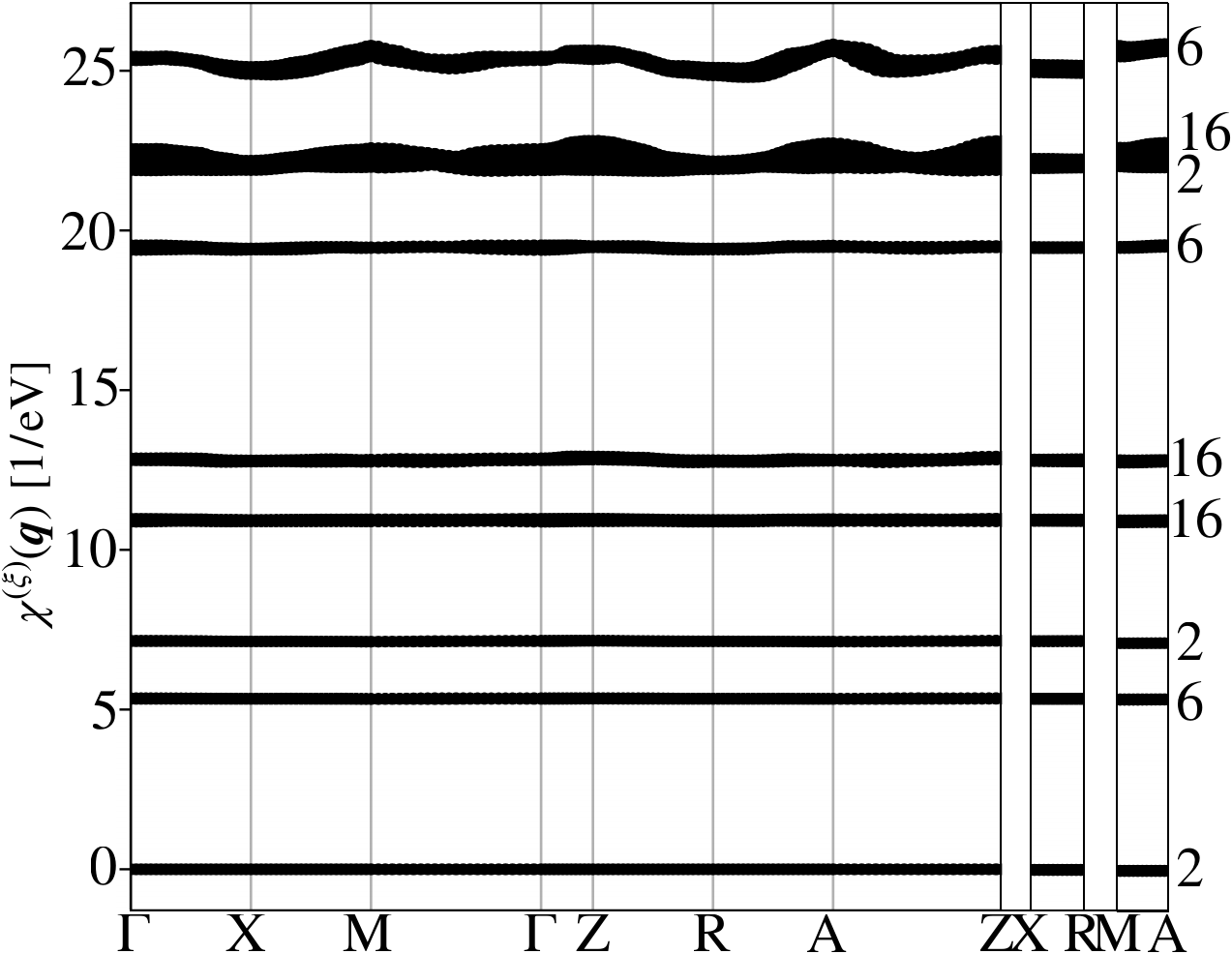}
    \caption{Eigenvalues $\chi^{(\xi)}(\bm{q})$ of the susceptibility for $T=0.01$~eV.
    The numbers on the right indicate the number of eigenmodes included in the group.}
    \label{fig:chi_eig}
\end{figure}

Figure~\ref{fig:chi_eig} shows the eigenvalues $\chi^{(\xi)}(\bm{q})$ on the $\bm{q}$ path that connects the symmetry points in the Brillouin zone.
There are 72 modes, which are composed of $6^2=36$ atomic degrees of freedom times two Ce atoms in a unit cell.
The structure of this graph can be understood in terms of the CEF states.
The six fluctuation modes in the top group are due to fluctuations within the ground-state doublet {\Gsevenone}, which has $2^2-1=3$ local degrees of freedom except for the charge fluctuation.
The second group with 16+2 fluctuation modes arises from mixing between the ground-state and first-excited doublets.
The third group is the fluctuations within the thermally excited {\Gsix} state.

\begin{figure}[t]
    \includegraphics[width=1.0\linewidth]{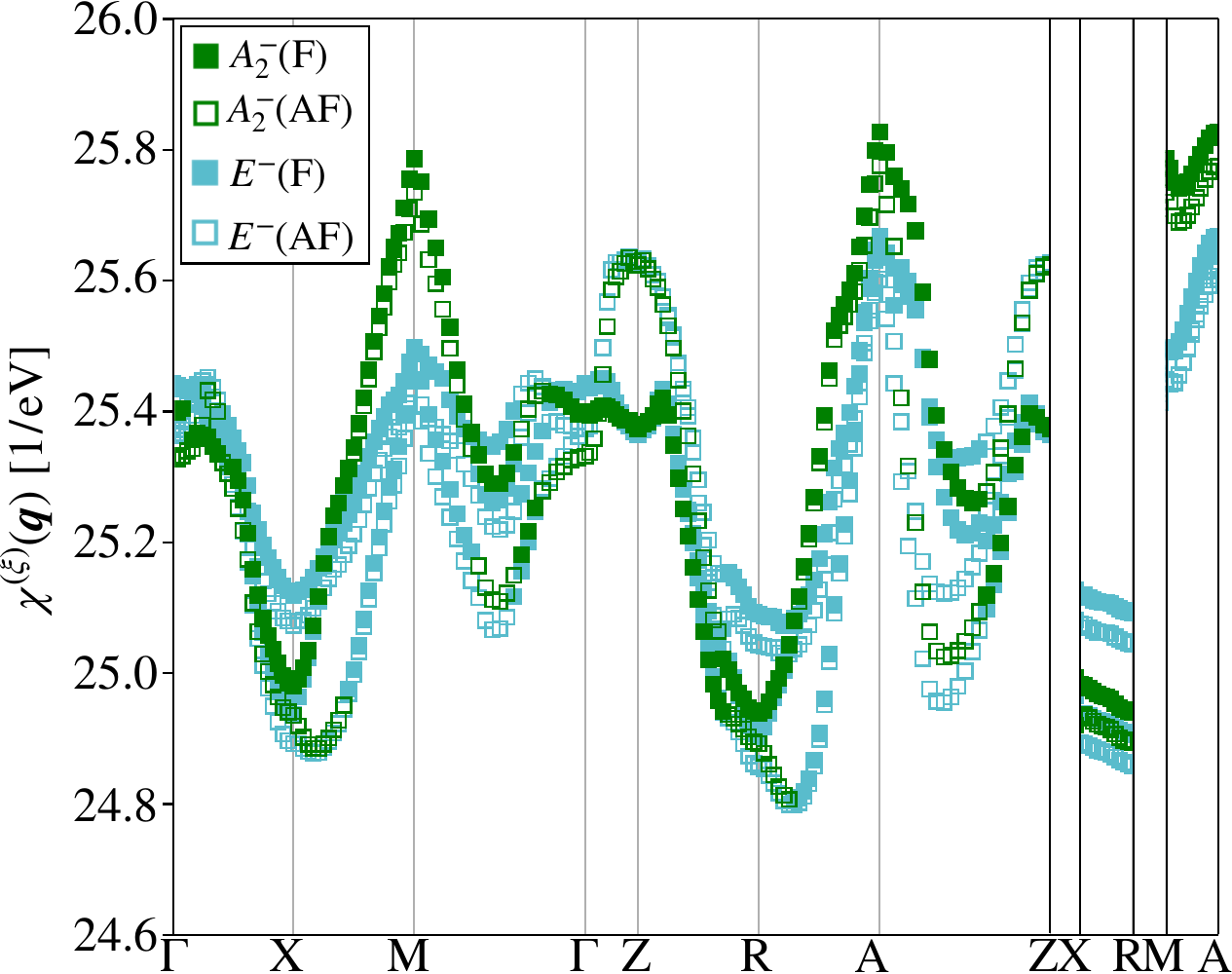}
    \caption{Eigenvalues $\chi^{(\xi)}(\bm{q})$ of the susceptibility that correspond to fluctuations within {\Gsevenone} doublet.}
    \label{fig:chi30-2}
\end{figure}

\begin{figure}[t]
    \begin{center}
        \includegraphics[width=0.4\linewidth]{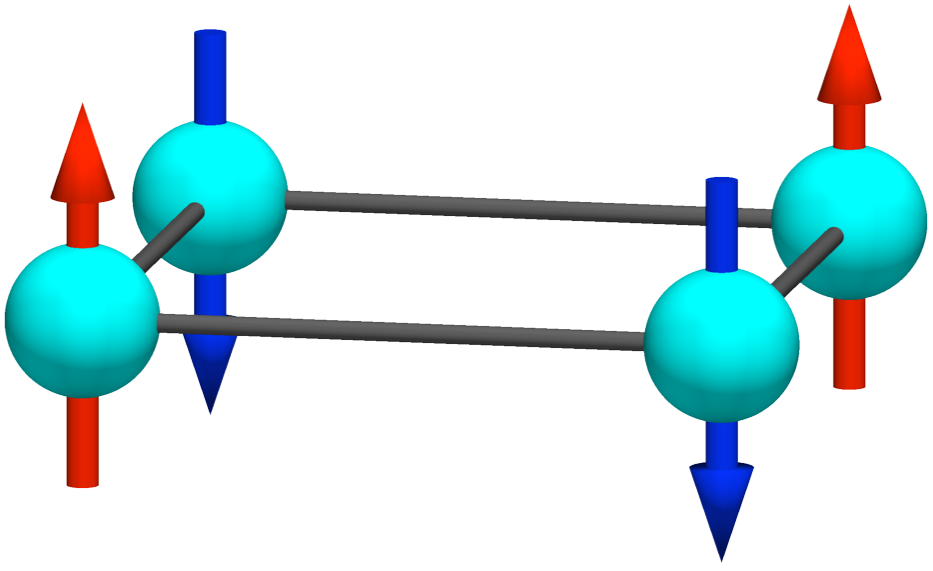}
    \end{center}
    \caption{Two-dimensional checkerboard type magnetic structure exhibiting the largest fluctuations in DFT+DMFT calculations.}
    \label{fig:mag_checker}
\end{figure}

We focus on the first group in Fig.~\ref{fig:chi_eig}.
Figure~\ref{fig:chi30-2} shows $\chi^{(\xi)}(\bm{q})$ with symmetry characters distinguished by colors and symbols.
The six fluctuation modes correspond to ferroic (F) or antiferroic (AF) configurations of the magnetic dipole ($M_x$, $M_y$, $M_z$) on the two Ce sites in a unit cell.
The leading instabilities occur in $M_z$ at $\bm{q}=\QA\equiv(1/2,1/2,1/2)$ and $\bm{q}=\QM\equiv(1/2,1/2,0)$ with nearly equal intensity, indicating a weak interlayer coupling between Ce sites.
Moreover, F and AF configurations are degenerate within numerical accuracy, suggesting that two Ce sites in a unit cell are effectively uncorrelated.
These results indicate a two-dimensional checkerboard magnetic structure shown in Fig.~\ref{fig:mag_checker}, with only weak interlayer coupling along the $c$ axis.

\begin{figure}[!t]
    \includegraphics[width=1.0\linewidth]{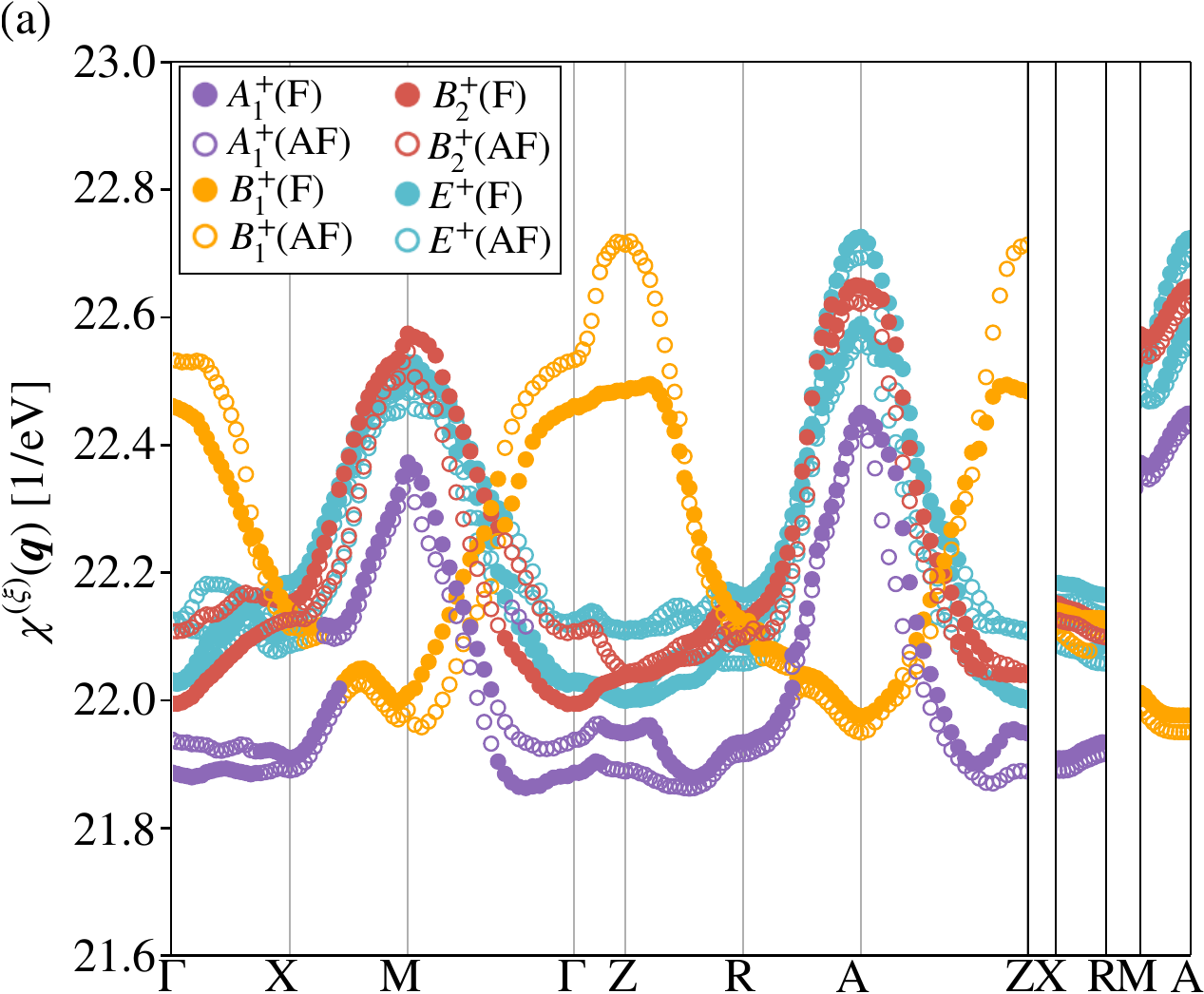}
    \includegraphics[width=1.0\linewidth]{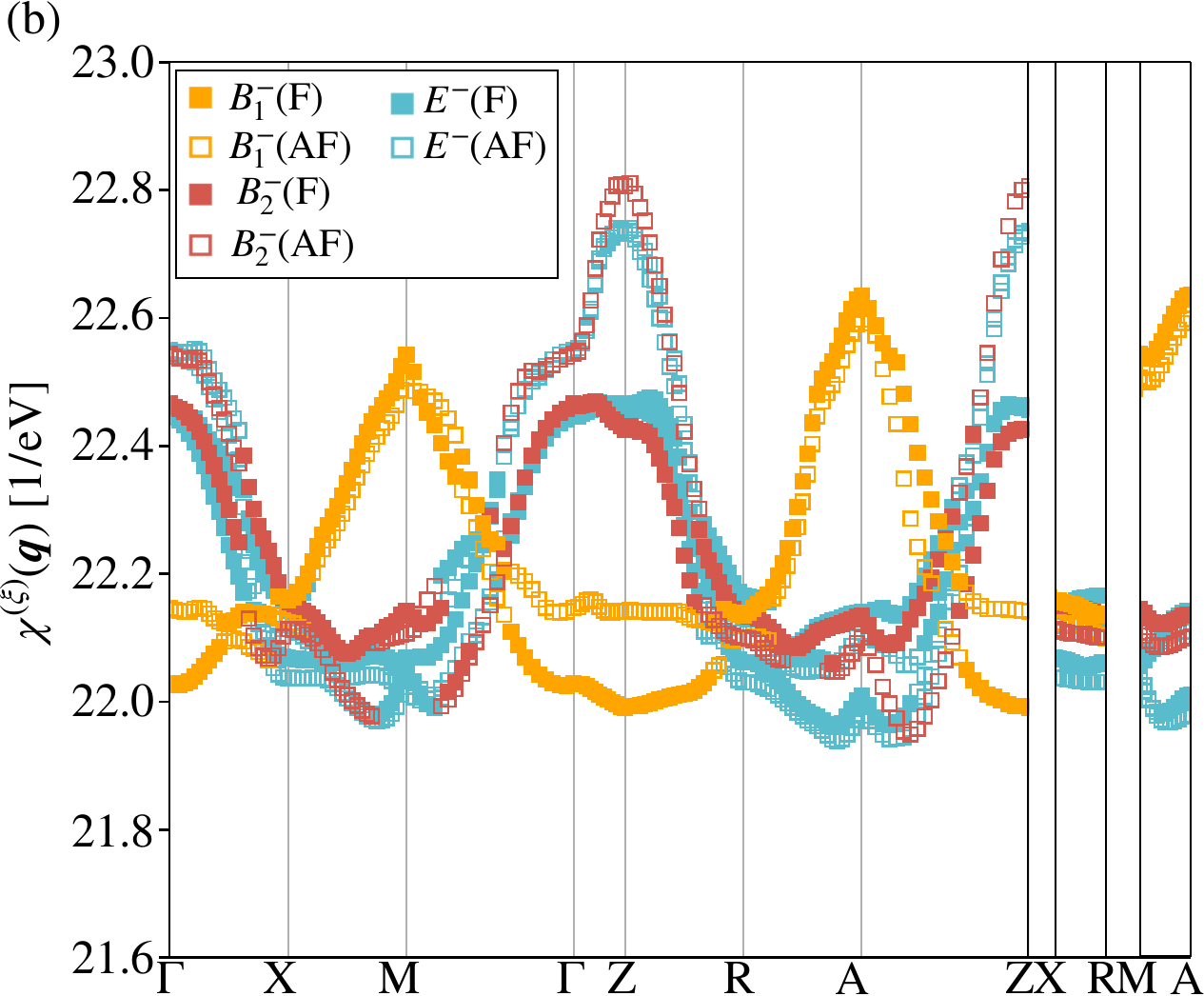}
    \caption{Eigenvalues $\chi^{(\xi)}(\bm{q})$ of the susceptibility within the second group that correspond to fluctuations between the {\Gsevenone} and {\Gsix} states. (a) Electric modes and (b) magnetic modes.}
    \label{fig:chiir_2nd}
\end{figure}

We next discuss the fluctuations in the second group that arise from the mixing of {\Gsevenone} and {\Gsix}.
Figure~\ref{fig:chiir_2nd} shows an enlarged plot of the eigenvalues $\chi^{(\xi)}(\bm{q})$ in the second group with additional symmetry labels.
Electric and magnetic modes are plotted separately in Fig.~\ref{fig:chiir_2nd}(a) and Fig.~\ref{fig:chiir_2nd}(b), respectively.
The largest fluctuation arises in the magnetic modes [Fig.~\ref{fig:chiir_2nd}(b)] with the octupole $M_{z(x^2-y^2)}$(AF) in $B_2^-$ symmetry at $\bm{q}=\QZ\equiv(0,0,1/2)$.
Magnetic dipoles $\{ M_x, M_y \}$(AF) follow this fluctuation.
In the electric modes [Fig.~\ref{fig:chiir_2nd}(a)], the leading fluctuations are the quadrupoles $Q_{x^2-y^2}$(AF) in $B_1^+$ representation at $\bm{q}=\QZ$ and $\{ Q_{yz}, Q_{zx} \}$ in $E^+$ representation at $\bm{q}=\QA$.

\subsection{Results for multipolar interactions}
\label{sec:interaction}

\begin{figure}[!t]
    \includegraphics[width=0.95\linewidth]{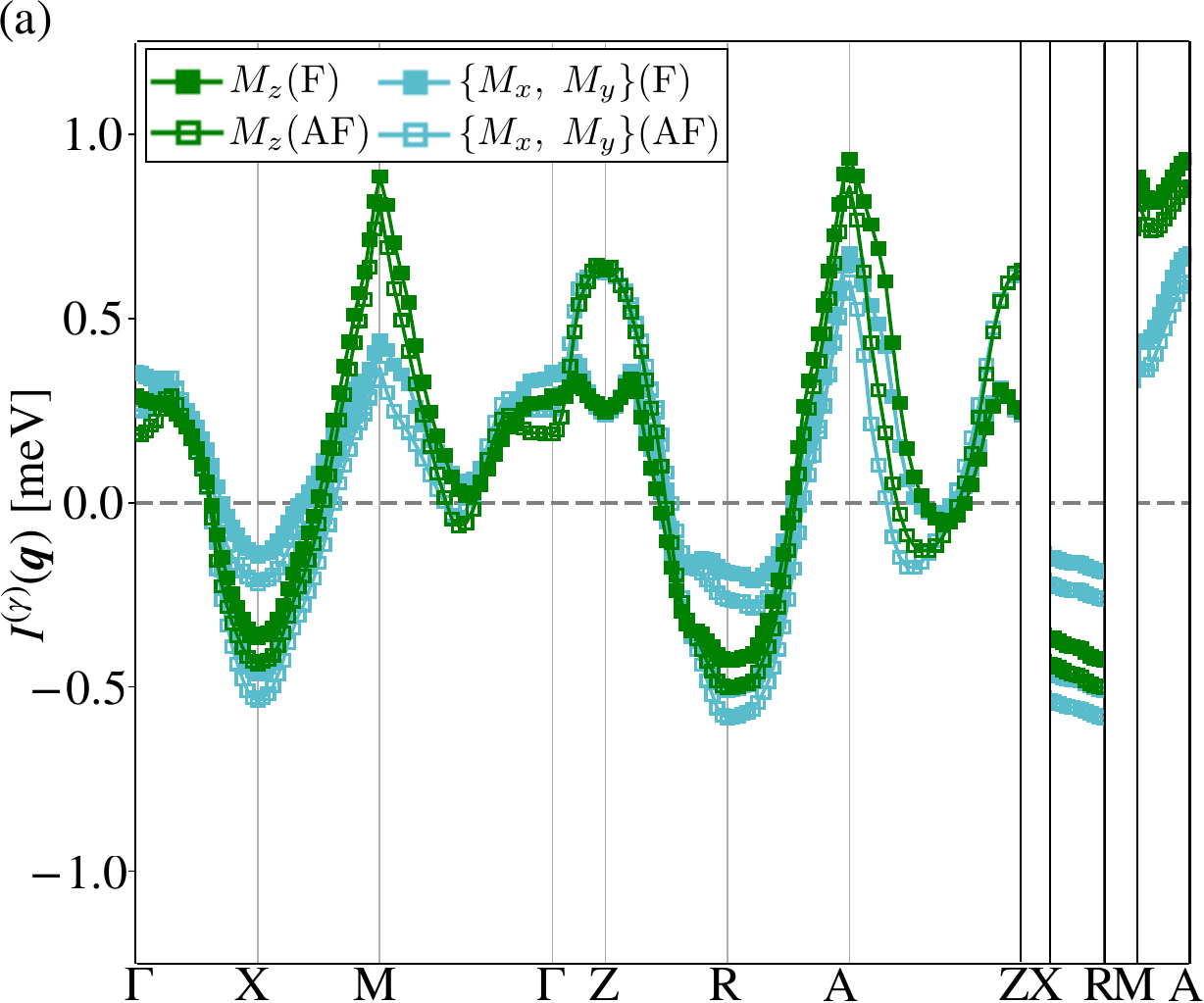}
    \includegraphics[width=0.95\linewidth]{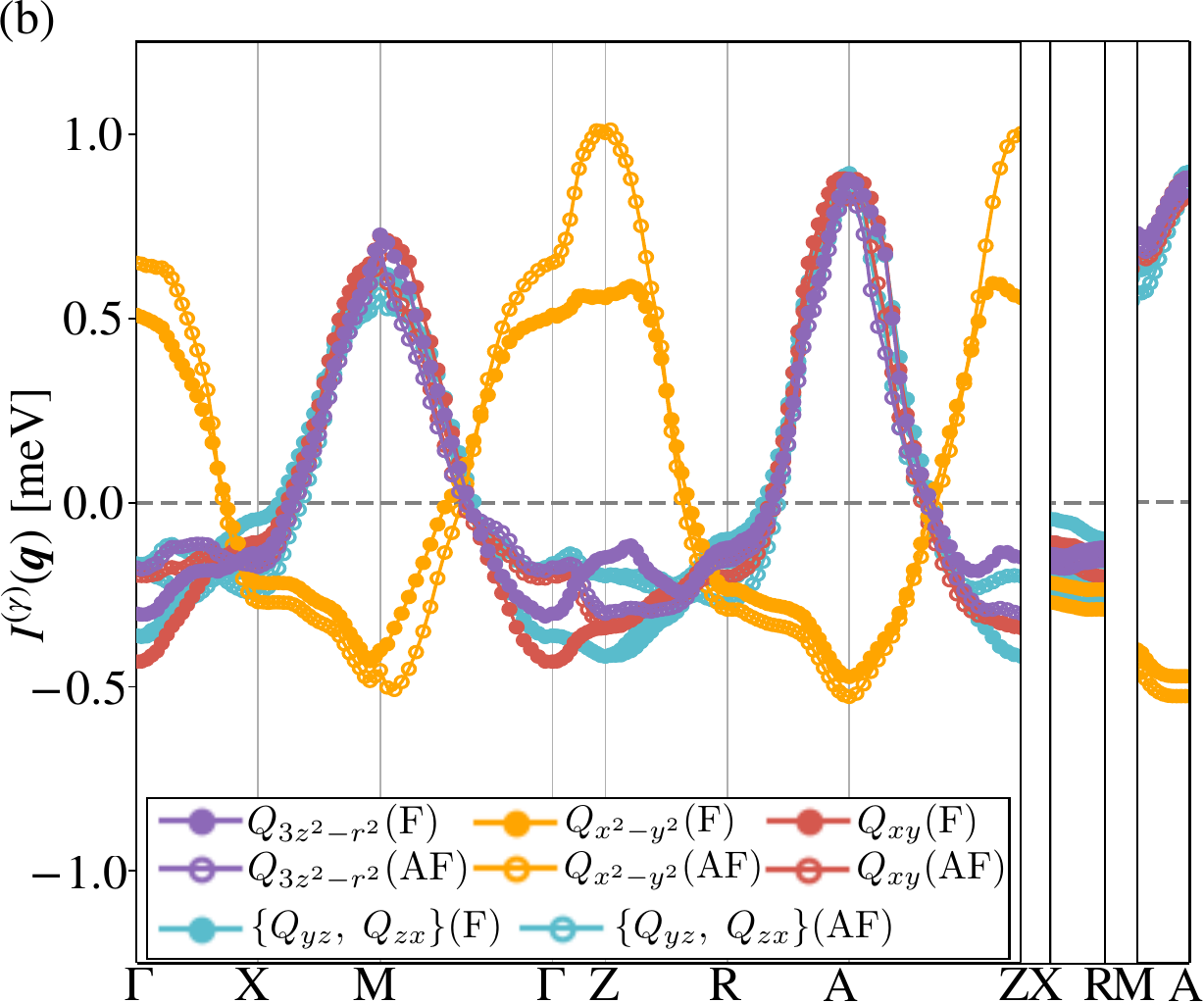}
    \includegraphics[width=0.95\linewidth]{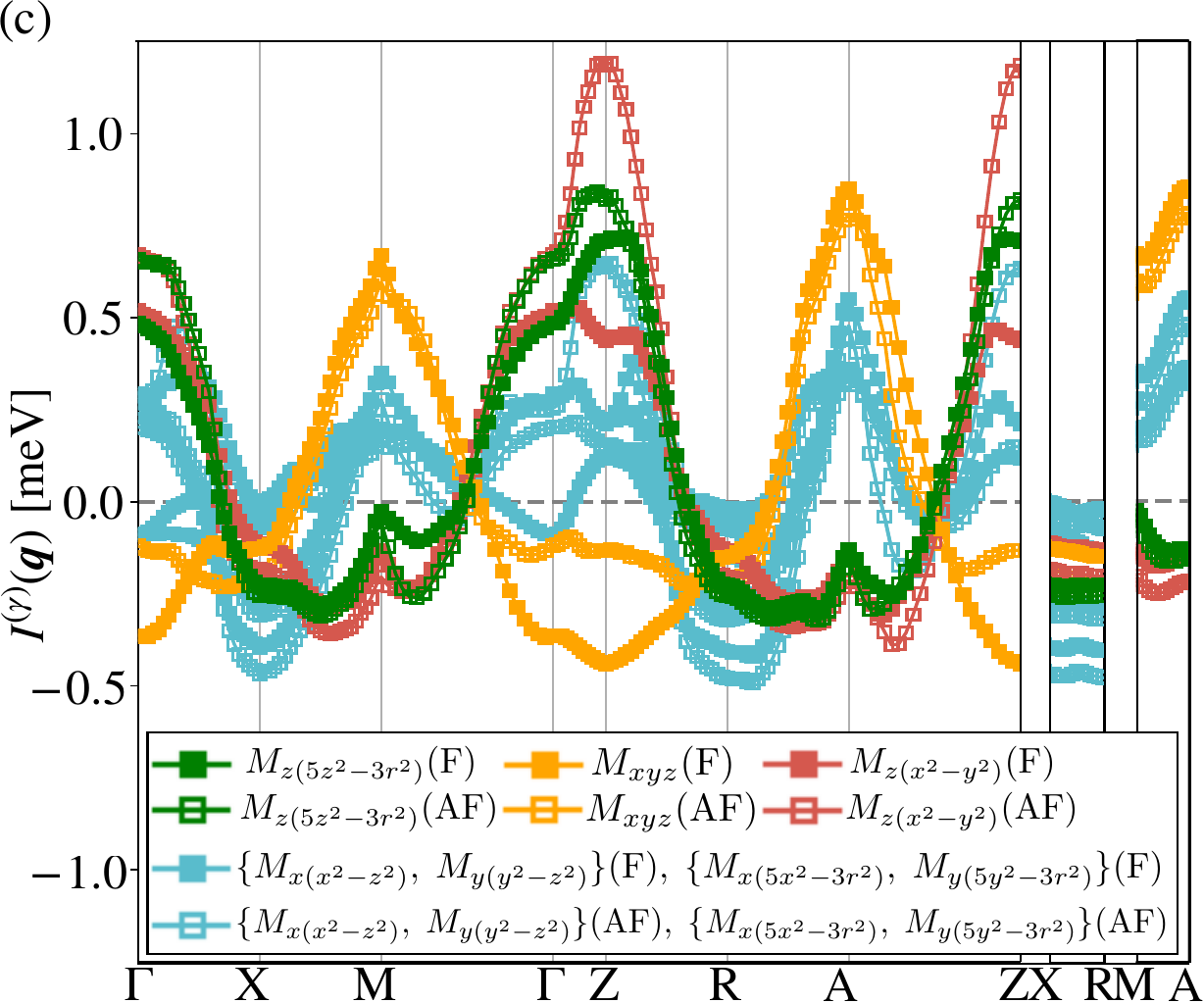}
    \caption{The effective multipolar interactions $I^{(\gamma)}(\bm{q})$ for (a) magnetic dipoles, (b) electric quadrupoles, and (c) magnetic octupoles.}
    \label{fig:Iq}
\end{figure}

We derive the effective interactions between multipoles.
This quantity allows us to access lower temperatures (Sec.~\ref{sec:temp-dep}) and to discuss the magnetic-field dependence of the transition temperature (Sec.~\ref{sec:anisotropy}).
The multipolar interactions are obtained by transforming $\hat{I}(\bm{q})$ into the SAMB representation.
For a given basis $\gamma$, the effective interaction is computed as
\begin{align}
    I^{(\gamma)}(\bm{q})
    &=\sum^{}_{ij}\sum^{}_{m_1,\cdots,m_4}
    \notag \\
    &\times z^{(\gamma)}_{i m_1m_2}(\bm{q}) I_{i m_1m_2, j m_3m_4}(\bm{q}) z^{(\gamma)}_{j m_3m_4}(\bm{q})^{\ast}.
\end{align}
Figure~\ref{fig:Iq}(a) shows interactions between the magnetic dipoles in the ground-state doublet {\Gsevenone}, whereas Figs.~\ref{fig:Iq}(b) and \ref{fig:Iq}(c) show the interactions between the electric quadrupoles and magnetic octupoles, which arise from the mixing between the ground-state and first-excited doublets.
The $\bm{q}$ dependence closely follows that in $\chi^{(\gamma)}(\bm{q})$.
The magnitude of the interaction is of the order of 1\,meV.
We remark that the interactions involving quadrupoles and octupoles are stronger than those between dipoles.
In particular, the electric quadrupole $Q_{x^2-y^2}(\textrm{AF})$ at $\bm{q}=\QZ$ and the magnetic octupole $M_{z(x^2-y^2)}(\textrm{AF}$) at $\bm{q}=\QZ$ exhibit especially strong interactions.
This implies that fluctuations of the induced higher-order multipoles are expected to play some significant role.

\subsection{Temperature dependence}
\label{sec:temp-dep}
So far, we presented results for a fixed temperature of $T=0.01$~eV. We now discuss the temperature dependence of the susceptibilities and determine the transition temperature.
As indicated in Eq.~\eqref{eq:chi_q_matrix}, $\hat{\chi}(\bm{q})$ consists of two contributions: the local susceptibility $\hat{\chi}_\mathrm{loc}$ and the effective intersite interaction $\hat{I}(\bm{q})$.
Previous work on \ce{CeB6} has shown that the temperature dependence is dominated by $\hat{\chi}_\mathrm{loc}$, allowing $\hat{I}(\bm{q})$ to be treated as effectively temperature independent~\cite{Otsuki2024-ceb6}.

\begin{figure}[t]
    \includegraphics[width=1.0\linewidth]{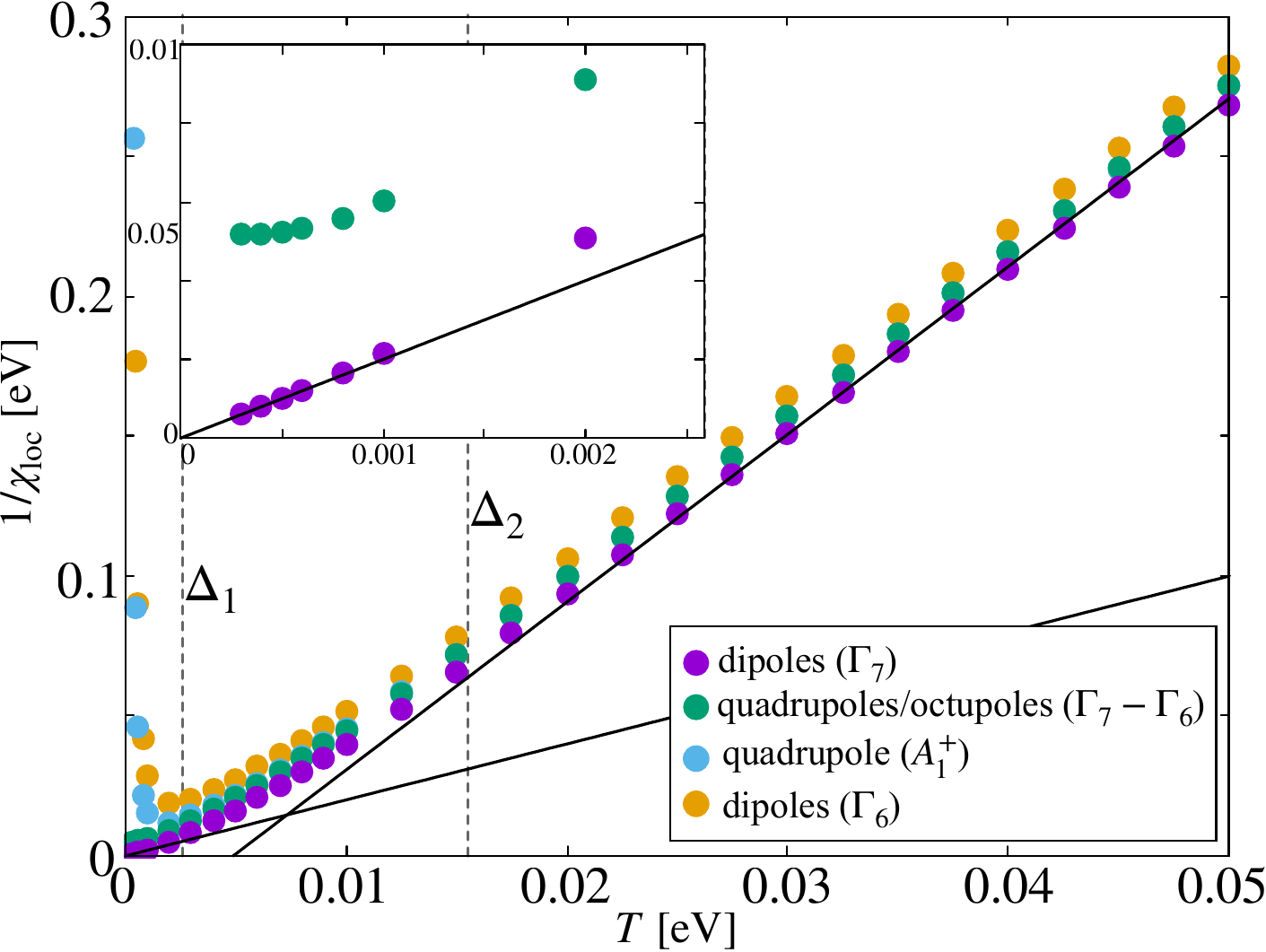}
    \caption{Temperature dependence of the inverse of the eigenvalues of $\hat{\chi}_\mathrm{loc}$. The solid lines show the Curie-Weiss law at high $T$ and the Curie law at low $T$. The vertical dashed lines indicate the energy of the CEF excited states in Fig.~\ref{fig:cef_split}(a).}
    \label{fig:chiloc_tdep}
\end{figure}

Figure~\ref{fig:chiloc_tdep} shows the temperature dependence of the eigenvalues $\chi_\mathrm{loc}^{(\xi)}$ of $\hat{\chi}_\mathrm{loc}$.
The leading eigenmodes are sixfold degenerate, corresponding to the magnetic dipoles within the CEF ground-state doublet at the two Ce sites in the unit cell.
At high temperatures $T>\Delta_2$, the behavior follows the Curie-Weiss law
\begin{align}
    \chi_\mathrm{loc}^{(\xi)} \approx \frac{C_\mathrm{high}}{T-\Theta},
\end{align}
with the Curie constant $C_\mathrm{high}=1/6$. In contrast, at low temperatures $T<\Delta_1$, the susceptibility follows the Curie law
\begin{align}
    \chi_\mathrm{loc}^{(\xi)} \approx \frac{C_\mathrm{low}}{T},
    \label{eq:chi-low-T}
\end{align}
with $C_\mathrm{low}=1/2$.
The lines in Fig.~\ref{fig:chiloc_tdep} indicate these asymptotic behaviors.

At high-symmetry $\bm{q}$ points, the multipole basis $\gamma$ can be chosen to coincide with the eigenmode $\xi$, since different irreducible representations do not mix.
In this case, the matrix equation \eqref{eq:chi_q_matrix} decouples into an equation for each eigenmode $\xi$ as
\begin{align}
    \chi^{(\xi)}(\bm{q}) = \frac{1}{(\chi_\mathrm{loc}^{(\xi)})^{-1} - I^{(\gamma)}(\bm{q})}.
\end{align}
Using Eq.~\eqref{eq:chi-low-T} and the value of $I^{(\gamma)}(\bm{q})$ evaluated at a given temperature, we can estimate the transition temperature from the divergence of $\chi^{(\xi)}(\bm{q})$.
For the leading fluctuation mode, namely, the magnetic dipole $M_z$ at $\bm{q}=\QA$, we obtain $I^{(\gamma)}(\bm{q})=0.93$~meV.
The transition temperature is estimated to be 5.4~K, which is an order of magnitude larger than the experimental value $T_0 \approx 0.5$~K.
Although DMFT commonly overestimates transition temperatures, the typical discrepancy is only a factor of 1.3--1.6~\cite{Otsuki2024-ceb6,cerh6ge4-itokazu}.
The much larger deviation in the present case may indicate the influence of the Kondo effect in {\cerhas}, as suggested in Refs.~\cite{Khim2021,Cristovam2024}.
In particular, the Hubbard-I approximation employed in this study neglects the Kondo effect and hence yields a transition temperature higher than a rigorous result within DMFT, especially for $U/W \lesssim 1$, where $U$ and $W$ denote the local Coulomb interaction and the bandwidth, respectively~\cite{Otsuki2019-SCL}.
The use of more advanced impurity solvers, such as continuous-time quantum Monte Carlo (CT-QMC)~\cite{Gull2011}, is expected to reduce this discrepancy.

In Fig.~\ref{fig:chiloc_tdep}, the susceptibilities of quadrupoles and octupoles saturate to a constant value at low temperatures. This behavior arises because these multipoles are generated through hybridization between the CEF ground state and the first-excited doublet, and therefore follow the van Vleck mechanism.
An exception is the $Q_{3z^2-r^2}$ quadrupole belonging to $A_{1}^{+}$ representation.
Its fluctuation is rapidly suppressed as the temperature is lowered below $T=\Delta_1$, similar to the fluctuations within the CEF excited doublet {\Gsix}.

\section{Anisotropy of $T$--$H$ phase diagram}
\label{sec:anisotropy}

The $T$--$H$ phase diagram in {\cerhas} is highly anisotropic as schematically shown in Figs.~\ref{fig:str_phase}(b) and \ref{fig:str_phase}(c): Phase~I is suppressed by the external magnetic field along the $c$ axis, while it is enhanced by the field in the $a$--$b$ plane.
Schmidt and Thalmeier attributed this anisotropy to the role of field-induced quadrupole $Q_{xy}$ under antiferromagnetic order of $\{ M_{x}, M_{y} \}$~\cite{Schmidt2024}. Their discussions are based on the assumption that $M_{x}$ and $Q_{xy}$ both have antiferroic intersite interactions of comparable magnitude.
Using the effective interactions presented in the previous section, we examine possible order parameters that can account for the anisotropy of the phase diagram.

\subsection{Phenomenology}

The enhancement of the transition temperature in a magnetic field has been investigated in the context of the quadrupolar ordering in \ce{CeB6}.
Following Ref.~\cite{Shiina1997}, we present a phenomenological description of the field-dependence of the transition temperature.
We show below that the transition temperature depends on fluctuations of the induced multipoles, even in the presence of the CEF splitting.

We consider the Landau free energy $\mathcal{F}(\phi_1, \phi_2, m)$ as a function of the primary order parameter $\phi_1$, the magnetic moment $m$ coupled to the external magnetic field, and the field-induced multipole $\phi_2$.
Near the critical temperature, $\mathcal{F}(\phi_1, \phi_2, m)$ can be expanded up to the fourth order as follows:
\begin{align}
    \mathcal{F}(\phi_1, \phi_2, m)
    &= F_0
    +\frac{1}{2} \chi_1^{-1}(T)\phi_1^2 + B\phi_1^4 + \frac{1}{2} \chi_2^{-1}(T)\phi_2^2 \notag \\
    &-Hm+\frac{1}{2} \chi_m^{-1}(T) m^2-gm\phi_1 \phi_2,
    \label{eq:F}
\end{align}
where $F_0$ denotes the contribution independent of $\phi_1$, $\phi_2$, and $m$.
The second and third terms describe a symmetry breaking for $\phi_1$.
The coefficient $\chi_1(T)$ is the susceptibility for $\phi_1$ in the paramagnetic state with $H=0$. This ensures the usual relation $\pdv*[2]{\mathcal{F}}{\phi_1}|_{\phi_1=0} = \chi_1^{-1}(T)$~\cite{Landau1996-stat}.
We neglected the fourth-order terms in $\phi_2$ and $m$, since we consider only the symmetry breaking of $\phi_1$.
The first term on the second line corresponds to the Zeeman energy, which induces a uniform magnetic moment $m$ along the field direction.
The last term describes the coupling among $\phi_1$, $\phi_2$, and $m$.
The coupling is allowed when the product $m \phi_1 \phi_2$ is invariant under the symmetry operation of the crystal.
The explicit combinations for $\phi_1$, $\phi_2$, and $m$ will be presented later.

Minimization of $\mathcal{F}$ with respect to $\phi_1$, $\phi_2$, and $m$ yields coupled equations that determine the thermodynamic values of $\phi_1$, $\phi_2$, and $m$. The transition temperature $T_1(H)$ is given by the temperature at which the symmetry-broken solution for $\phi_1$ vanishes. In the lowest order of $g$, this condition reads
\begin{align}
    \label{eq:t1_rel}
      \left[\chi_{1}^{-1}(T)-\frac{g^2H^2}{\chi_{2}^{-1}(T) \chi_m^{-2}(T)}\right]_{T=T_1(H)}=0.
\end{align}
An explicit form of $T_1(H)$ depends on the form of $\chi_1(T)$ and $\chi_2(T)$. We consider two cases below.

We first discuss the case of \ce{CeB6}, following Ref.~\cite{Shiina1997}.
The primary order parameter $\phi_1$ and the induced moment $\phi_2$ correspond to the electric quadrupole and the magnetic octupole, respectively.
Both multipoles are active in the CEF ground-state quartet.
In this case, the corresponding susceptibilities follow the Curie-Weiss law represented by
\begin{align}
    \label{eq:t-chi1}
    &\chi_1^{-1}(T)=\frac{T}{C}-I_1, \\
    \label{eq:t-chi2}
    &\chi_2^{-1}(T)=\frac{T}{C}-I_2.
\end{align}
Here, $\chi_1$ and $\chi_2$ differ in the intersite interactions $I_1$ and $I_2$, whereas the Curie constant $C$ is common to both.
By definition, the primary order parameter has a larger interaction, $I_1 > I_2$.
Substituting these expressions into Eq.~\eqref{eq:t1_rel} and solving the resulting quadratic equation for $T_1(H)$, we obtain, to the lowest-order in $g$,
\begin{align}
        \label{eq:th/t}
      \frac{T_1(H)}{T_1(0)} \simeq 1 +\left[\frac{1}{1-I_2/I_1}\right]
      g^2 C^2 \chi_m\big(T_1(0)\big)^2 h^2,
\end{align}
where $T_1(0)=CI_1$.
We introduced the dimensionless magnetic field $h = H/T_1(0)$.
The coefficient of the $h^2$ term is always positive, meaning that the induced moment enhances the $T_1(H)$ under the magnetic field.
The magnitude of the enhancement depends on the ratio $I_2/I_1$. To highlight the influence of the induced multipole, we introduce an enhancement factor defined by
\begin{align}
    \label{eq:enhance1}
      \alpha_1=\frac{1}{1- I_2 / I_1}.
\end{align}
In \ce{CeB6}, the antiferro-quadrupolar phase of $Q_{xy}$ type (phase~II) is enhanced under the magnetic field through the induced octupole of $M_{xyz}$ type~\cite{Shiina1997}.
This behavior is ascribed to the relation $I_{2}/I_{1} \approx 1$, which has been proven analytically~\cite{Shiba1999} and demonstrated numerically~\cite{yamada_ceb6,Otsuki2024-ceb6}.

In the case of {\cerhas}, the situation differs due to the CEF splitting.
Suppose that the primary order parameter $\phi_1$ is the magnetic dipole within the CEF ground-state doublet, then $\chi_1(T)$ follows the Curie-Weiss law as in Eq.~\eqref{eq:t-chi1}.
In contrast, the induced quadrupole arises from mixing between the ground-state and excited doublets.
As a result, $\chi_2(T)$ does not follow the Curie-Weiss law but instead saturates to a constant at low temperatures, as shown in Fig.~\ref{fig:chiloc_tdep}.
Replacing $\chi_2(T)$ with $\chi_2(0)$ in Eq.~\eqref{eq:t1_rel} and using Eq.~\eqref{eq:t-chi1}, we obtain
\begin{align}
    \frac{T_1(H)}{T_1(0)}&=1+I_1\chi_2(0) g^2 C^2 \chi_m\big(T_1(0)\big)^2 h^2.
\end{align}
The enhancement factor $\alpha_2$ is therefore given by
\begin{align}
    \label{eq:enhance2}
    \alpha_2=I_1\chi_2(0).
\end{align}
In the next section, we discuss the order parameter of phase~I in {\cerhas} based on the value of $\alpha_2$.

\subsection{Application of DFT+DMFT results}

\begin{table*}[!t]
      \centering
      \caption{Candidates of the magnetic dipole order parameter $\phi_1$ in phase~I and the corresponding field-induced electric quadrupole $\phi_2$. The column ``F/AF" indicates ferroic and antiferroic configurations within a unit cell, where ``F/AF" in an entry indicates that F and AF are degenerate. The enhancement factor $\alpha_2$ is defined in Eq.~\eqref{eq:enhance2}.
      $M_{x+y}=M_x+M_y$ and $Q_{(x+y)z}=Q_{yz}+Q_{zx}$ represent linear combinations of the multipoles belonging to $E$ representation.}
      \label{tab:enhance_dipole}
      \begin{tabular}{cccc|cccc|cccc}\hline
            $\phi_1$  & F/AF & $\bm{q}$ & $T_1$ [K] & Direction of $m$ & $\phi_2$       & $I_2/I_1$ & $\alpha_2$ & Direction of $m$ & $\phi_2$ & $I_2/I_1$ & $\alpha_2$ \\\hline
            $M_z$     & F/AF   & $\QA$    & $5.4$     & [001]            & $Q_{3z^2-r^2}$ & $0.94$    & $0$        & [100]             & $Q_{zx}$           & $0.96$  & $0.22$     \\
                      & F/AF   & $\QM$    & $5.1$     &                  &                & $0.82$    & $0$        &                   &                    & $0.71$  & $0.20$     \\
                      & AF   & $\QZ$    & $3.7$     &                  &                & $-0.48$   & $0$        &                   &                    & $-0.32$ & $0.12$     \\\hline
            $M_y$     & F/AF   & $\QA$    & $3.9$     & [001]            & $Q_{yz}$       & $1.3$     & $0.16$     & [100]             & $Q_{xy}$           & $1.3$   & $0.16$     \\
                      & AF   & $\QZ$    & $3.6$     &                  &                & $-0.32$   & $0.12$     &                   &                    & $-0.53$ & $0.11$     \\\hline
            $M_{x+y}$ & F/AF   & $\QA$    & $3.9$     & [001]            & $Q_{(x+y)z}$   & $1.5$     & $0.17$     & $[1\bar{1}0]$     & $Q_{x^2-y^2}$      & $-0.70$ & $0.12$     \\
                      & AF   & $\QZ$    & $3.6$     &                  &                & $-0.31$   & $0.12$     &                   &                    & $1.6$   & $0.15$     \\\hline
      \end{tabular}
\end{table*}

\begin{figure*}[!t]
    \includegraphics[width=0.7\linewidth]{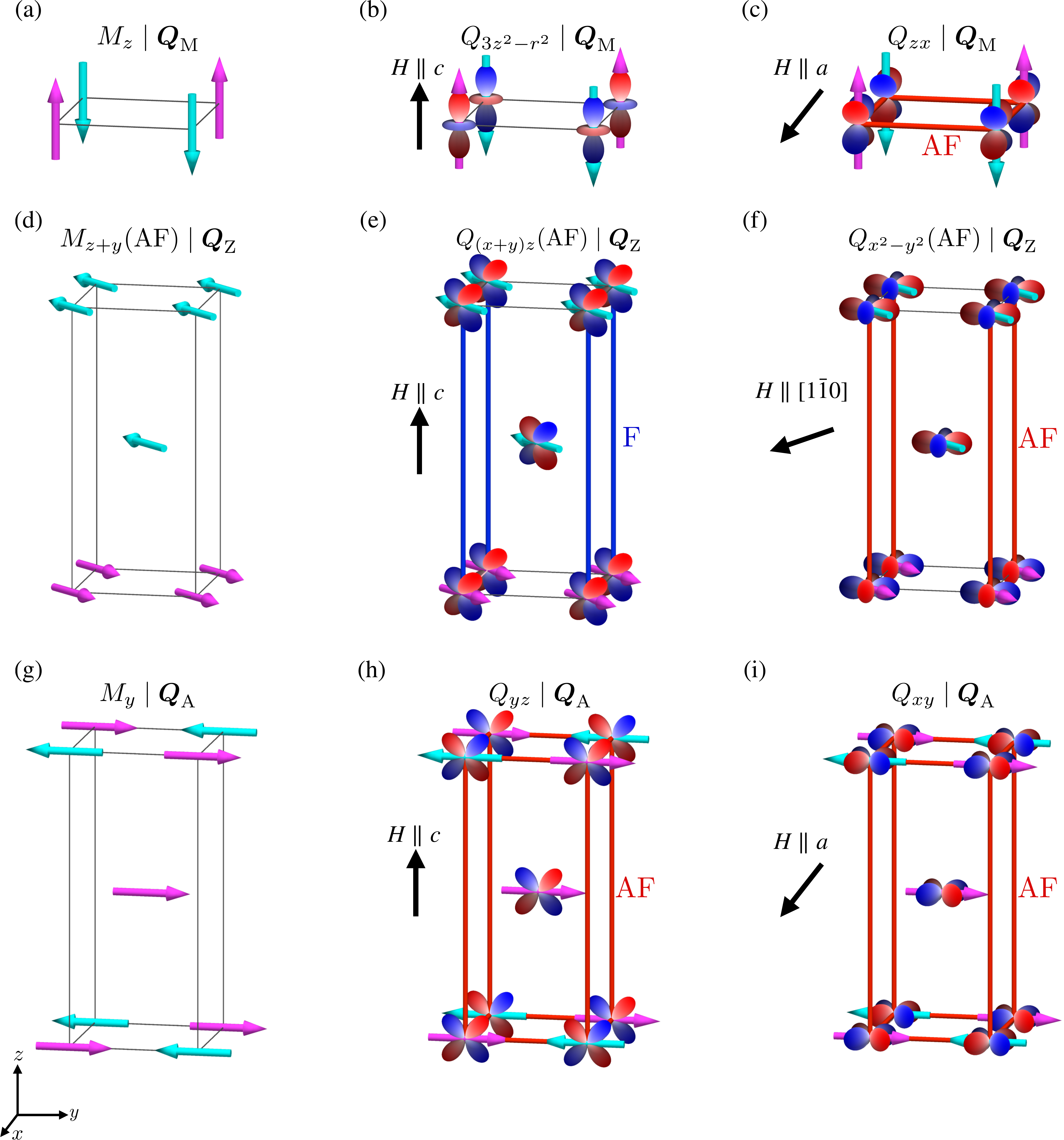}
    \caption{Candidate magnetic structures and the induced quadrupoles under magnetic fields $H \parallel c$ and $H \perp c$. (a)--(c) $M_z$ order with $\bm{q}=\QM$, (d)--(f) $M_{x+y}(\textrm{AF})$ order with $\bm{q}=\QZ$, and (g)--(i) $M_y$ order with $\bm{q}=\QA$. The red and blue bonds represent the antiferroic and ferroic quadrupolar correlations, respectively. The labels such as $M_z\,|\,\QM$ indicate the ordered state of $M_z$ at $\bm{q}=\QM$.}
    \label{fig:di-quad}
\end{figure*}

We now incorporate our DMFT results into the above phenomenological description to discuss the experimentally determined anisotropic $T$--$H$ phase diagram.
In this analysis, we assume that the primary order parameter $\phi_1$ corresponds to the magnetic dipole within the CEF ground-state doublet, and then the induced moment $\phi_2$ is the electric quadrupole.
The combinations of $\phi_1$, $\phi_2$, and $m$ that are allowed from the symmetry are summarized in Table~\ref{tab:enhance_dipole}.
Here, $\phi_1$ and $\phi_2$ share the same ordering vector $\bm{q}$, whereas $m$ is uniform.
Figure~\ref{fig:di-quad} illustrates the configurations of three candidates for the magnetic structure and the induced quadrupoles under magnetic fields $H \parallel c$ and $H \perp c$.

For the magnetic moment $M_z$ along the $c$ axis, a magnetic field $H \parallel c$ induces the quadrupole of $Q_{3z^2-r^2}$, whereas a field $H \parallel a$ induces $Q_{zx}$.
This follows from the fact that the product $\phi_1 \phi_2 m$, corresponding to $M_z(\bm{q}) Q_{3z^2-r^2}(-\bm{q}) M_z(\bm{0})$ or $M_z(\bm{q}) Q_{zx}(-\bm{q}) M_x(\bm{0})$, is invariant under spatial inversion, translation, and time-reversal operations.
For the in-plane magnetic moment $M_y$, the magnetic field $H \parallel c$ induces $Q_{yz}$, while the field $H \parallel a$ induces $Q_{xy}$.
For the magnetic dipole oriented along the $[110]$ direction, denoted by $M_{x+y}$, the induced quadrupole is rotated by 45$^{\circ}$ and the in-plane field induces $Q_{x^2-y^2}$.

From the DMFT results in Sec.~\ref{sec:multi-chi}, we evaluated the magnetic transition temperatures $T_1$, the ratio of the intersite interactions $I_2/I_1$, and the enhancement factor $\alpha_2$ defined in Eq.~\eqref{eq:enhance2}.
The results are summarized in Table~\ref{tab:enhance_dipole}.
In this analysis, we selected specific $\bm{q}$ points that yield relatively large transition temperatures.
We note that $\alpha_2=0$ in the case with the induced quadrupole $Q_{3z^2-r^2}$ because its local fluctuation is suppressed at low temperatures, which can be confirmed in Fig.~\ref{fig:chiloc_tdep}.
For the other types of induced quadrupoles, $\alpha_2$ is finite since $\chi_2(0)$ remains finite owing to the van Vleck contribution.
For positive values of $I_2/I_1$, the enhancement factor $\alpha_2$ tends to become large, reflecting the growth of the fluctuation $\chi_2(0)$ of the induced quadrupole.

In the experiments, the transition temperature is suppressed for $H \parallel c$ and enhanced for $H \perp c$ as schematically shown in Fig.~\ref{fig:str_phase}. This behavior can be explained by a small enhancement factor $\alpha_2$ for $H \parallel c$ and a large $\alpha_2$ for $H \perp c$.
Two cases are consistent with the experimental anisotropy.
The first case is a primary magnetic dipole $M_z$ with $\bm{q}=\QA$ or $\QM$, which exhibits the largest fluctuations as shown in Fig.~\ref{fig:chi30-2}.
Figures~\ref{fig:di-quad}(a)--(c) show this magnetic structure and the corresponding field-induced quadrupoles under $H \parallel c$ and $H \perp c$.
For $H \parallel c$, no enhancement of the transition temperature is expected, because the fluctuations of the induced quadrupole $Q_{3z^2-r^2}$ belonging to $A_1^+$ representation are suppressed with decreasing temperature as demonstrated in Fig.~\ref{fig:chiloc_tdep}.
In contrast, for $H \perp c$, the transition temperature can increase due to the enhanced fluctuations of the induced quadrupole $Q_{zx}$, which are strengthened by the antiferroic interactions $I_2>0$.

\begin{figure}[t]
    \includegraphics[width=0.3\linewidth]{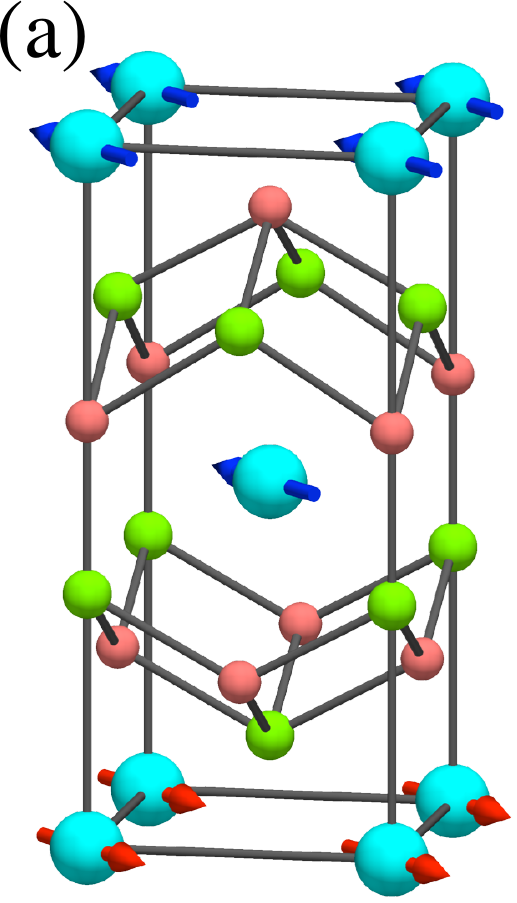}
    \includegraphics[width=0.3\linewidth]{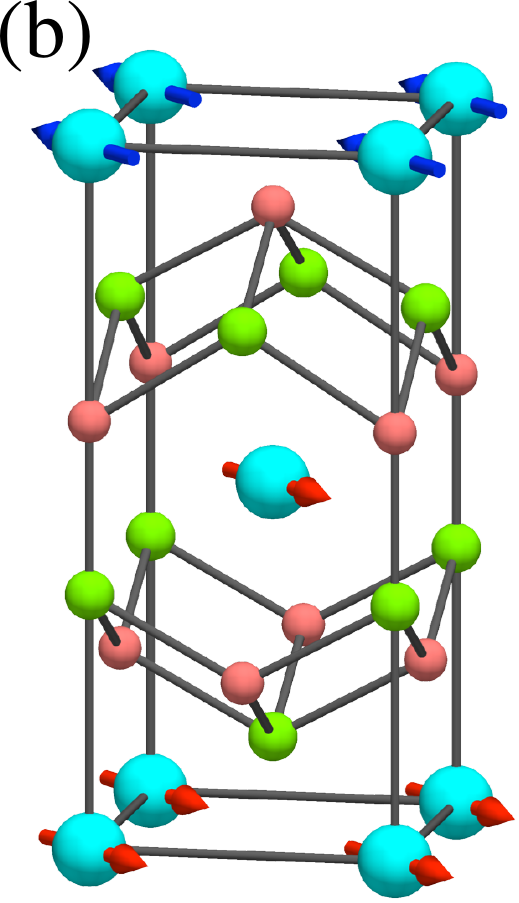}
    \caption{The magnetic structure for (a) $M_{x+y}(\textrm{AF})$ and (b) $M_{x+y}(\textrm{F})$ with $\bm{q}=\QZ$.}
    \label{fig:m110_z}
\end{figure}

The other case is the in-plane dipole ordering $M_{x+y}(\textrm{AF})$ with $\bm{q}=\QZ$, illustrated in Fig.~\ref{fig:di-quad}(d).
Under $H \parallel c$, the interaction between the induced quadrupole $Q_{(x+y)z}$ is ferroic, $I_2<0$, leading to a suppression of its fluctuations [Fig.~\ref{fig:di-quad}(e)]. In contrast, for $H \perp c$, the induced quadrupole $Q_{x^2-y^2}$ exhibits large (even larger than that for $M_{x+y}$) antiferroic interactions $I_2>0$, which enhance its fluctuations [Fig.~\ref{fig:di-quad}(f)].
The difference in the effective interaction of the induced quadrupoles, therefore, accounts for the anisotropy observed in the phase diagram.
We note that the configuration $M_{x+y}(\textrm{AF})$ differs from $M_{x+y}(\textrm{F})$. They are shown in Fig.~\ref{fig:m110_z}. These two magnetic structures are inequivalent because no mirror symmetry in the $a$--$b$ plane passes through the Ce site.
We also remark that the case with $M_x$ and $M_y$ moments leads to an anisotropy that is inconsistent with experiments since the effective interactions between the induced quadrupoles $Q_{xy}$ are ferroic.

Finally, a comment on the checkerboard magnetic structure with an in-plane moment is in order.
As an example, we consider the $M_y$ order with $\bm{q}=\QA$, which is illustrated in Fig.~\ref{fig:di-quad}(g).
The checkerboard magnetic structure of $M_y$ is proposed in Ref.~\cite{Schmidt2024}.
The induced quadrupole is $Q_{yz}$ for $H \parallel c$ and $Q_{xy}$ for $H \perp c$, as shown in Figs.~\ref{fig:di-quad}(h) and \ref{fig:di-quad}(i), respectively.
Table~\ref{tab:enhance_dipole} shows that both induced quadrupoles are subject to antiferroic interactions of the same magnitude.
Therefore, the anisotropy of the experimentally observed $T$--$H$ phase diagram is not expected in this case.

\section{Discussion}
\label{sec:Discussion}

\subsection{Magnetic ordering scenario for phase~I}

Based on the susceptibilities in Sec.~\ref{sec:multi-chi} and the expected anisotropy of the transition temperature in Sec.~\ref{sec:anisotropy}, we propose two candidates for the order parameter in phase~I:
The primary candidate is
the two-dimensional checkerboard configuration of $M_z$ as illustrated in Fig.~\ref{fig:mag_checker}, and the subleading one is the in-plane dipole ordering as illustrated in Fig.~\ref{fig:m110_z}.
We now focus on the former.
This configuration is consistent with the inelastic neutron scattering experiments, which report intensities at $\bm{q}=\QM$~\cite{Chen2024-neutron}. This translation vector can be accounted for by nesting of the quasi-two-dimensional Fermi surface in DFT calculations with localized $4f$ electrons.
Our results support antiferromagnetic ordering as proposed by $\mu$SR~\cite{Khim2025-musr}.

Upon the onset of this magnetic structure, the fourfold rotational symmetry is broken.
As a consequence, the maximal magnetic space group in the ordered phase becomes $C_{P}m^{\prime}ma$ (No.~67.13.589 in Opechowski-Guccione notation)~\cite{perez2015symmetry,Litvin2013}.
In particular, the related magnetic point group is $mmm1^{\prime}$, i.e., a gray magnetic point group that still contains time-reversal symmetry.
Importantly, this implies that all spatially averaged ($\bm{q}=\bm{0}$) time-reversal-odd quantities vanish, even though finite-$\bm{q}$ magnetic order is present.

\begin{figure}[t]
    \includegraphics[width=1.0\linewidth]{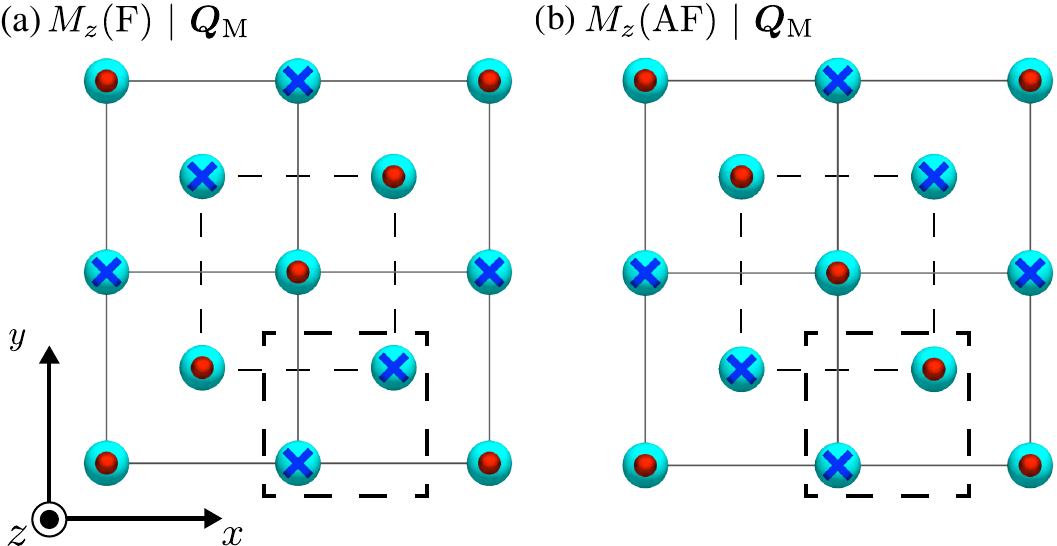}
    \caption{Checkerboard-type magnetic structures at $\bm{q}=\QM$, as viewed from the $z$ direction, with (a) ferroic and (b) antiferroic configurations in the unit cell (indicated by the enclosed thick dashed lines).
    The red dots and blue crosses indicate $+M_{z}$ and $-M_{z}$, respectively.
    Sites connected by thin solid (dashed) lines lie in the $z=0$ ($z=1/2$) plane.
    The labels follow the notation in Fig.~\ref{fig:di-quad}.}
    \label{fig:cheker_M+-}
\end{figure}

There are two degenerate configurations, $M_z(\mathrm{F})$ and $M_z(\mathrm{AF})$, as illustrated in Fig.~\ref{fig:cheker_M+-}.
From a symmetry point of view, the reduction of the magnetic point group from $4/mmm1^{\prime}$ to $mmm1^{\prime}$ allows an additional electric quadrupole component in the magnetically ordered phase~\cite{yatsushiroMultipoleClassification1222021}.
Since the present magnetic structure has mirror symmetries perpendicular to the $[110]$ or $[1\bar{1}0]$ directions, the $xy$-type electric quadrupole $Q_{xy}$ (in the original crystal coordinate system) is permitted.
Here, we introduce a domain parameter $\eta$ as the intensity difference: $\eta\equiv|M_{z}(\mathrm{F})|^{2}-|M_{z}(\mathrm{AF})|^{2}$.
This domain parameter transforms as $Q_{xy}$ and therefore couples linearly to the shear strain $\varepsilon_{xy}$.
Consequently, a symmetry-allowed coupling term of the form $\lambda\varepsilon_{xy}\eta$ appears in the Landau free energy, implying that uniaxial strain applied along the $[110]$ direction lifts the degeneracy between the two domains.
Therefore, when the system is cooled through $T_{0}$ under such strain, domain selection occurs in the ordered state, enabling statistical control of the checkerboard domains.

\subsection{Quadrupolar ordering scenario for phase~I}

We consider here a quadrupolar ordering scenario that becomes feasible when the CEF excited doublet participates in the low-energy physics, for instance, by the Kondo effect.
Once a pseudo-quartet is formed, the susceptibilities arising from hybridization between the CEF ground state {\Gsevenone} and the excited state {\Gsix} acquire significance.
We therefore examine the quadrupole scenario based on the DFT+DMFT results shown in Fig.~\ref{fig:chiir_2nd}.

Hafner \textit{et al.} proposed a quadrupolar density wave of $Q_{3z^2-r^2}$ with an incommensurate translation vector $\bm{q}=(q_x, 0, 0)$~\cite{Hafner2022}.
This state corresponds to the $A_{1}^+$ representation in our notation. In Fig.~\ref{fig:chiir_2nd}(a), however, the $A_{1}^+$ fluctuation is not enhanced along the $\Gamma$--$\mathrm{X}$ line relative to other quadrupole modes. Furthermore, the local fluctuations of the $A_{1}^+$ quadrupole rapidly vanish for $T<\Delta_1$, indicating that the incommensurate QDW is not favored within our localized $4f$ model.
On the other hand, Harima proposed antiferro-quadrupolar order of $Q_{xy}$~\cite{Harima2023}, which corresponds to $B_{2}^{+}$(AF) at $\bm{q}=\QG\equiv(0,0,0)$.
In Fig.~\ref{fig:chiir_2nd}(a), this quadrupole mode exhibits pronounced fluctuations not at $\bm{q}=\QG$ but rather at $\bm{q}=\QA$.
In our results, fluctuations at $\bm{q}=\QG$ remain generally weak, implying antiferroic correlations between Ce sites along the $c$ axis.

\begin{figure}[t]
    \includegraphics[width=1.0\linewidth]{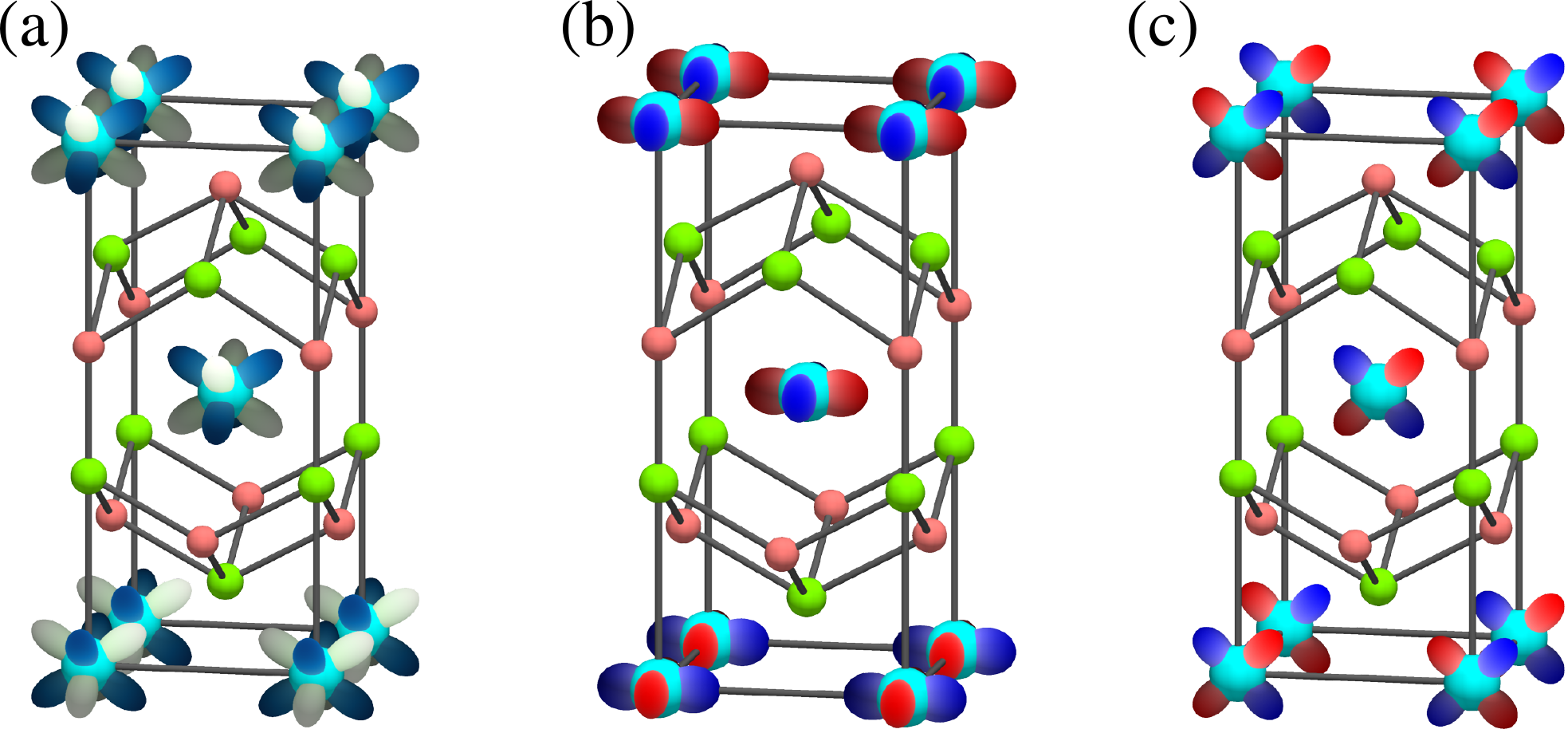}
    \caption{Configurations corresponding to large fluctuations arising by hybridization between {\Gsevenone} and {\Gsix}. (a) The magnetic octupole $M_{z(x^2-y^2)}$(AF) at $\bm{q}=\QZ$, (b) electric quadrupole $Q_{x^2-y^2}$(AF) at $\bm{q}=\QZ$, and (c) electric quadrupole $\{ Q_{yz}, Q_{zx} \}$ at $\bm{q}=\QA$.
    The different colors stand for the sign of the magnetic density in (a) and the sign of the charge density in (b) and (c).}
    \label{fig:phase1_cand}
\end{figure}

From the DFT+DMFT results in Fig.~\ref{fig:chiir_2nd},
the leading instability in the pseudo-quartet system is the magnetic octupole $M_{z(x^2-y^2)}$(AF) at $\bm{q}=\QZ$, followed by the electric quadrupoles $Q_{x^2-y^2}$(AF) at $\bm{q}=\QZ$ and $\{ Q_{yz}, Q_{zx} \}$ at $\bm{q}=\QA$.
Figure~\ref{fig:phase1_cand} illustrates the magnetic and electric configurations corresponding to these fluctuations. Applying the method described in Sec.~\ref{sec:anisotropy}, we investigated how the magnetic field affects the transition temperature. We found, however, that none of these candidates yields the anisotropy consistent with the experiments. Further details are presented in Appendix~\ref{app:anisotropy}.

\subsection{Magnetic structure in SC1+AFM phase}

\begin{figure}[t]
    \includegraphics[width=0.63\linewidth]{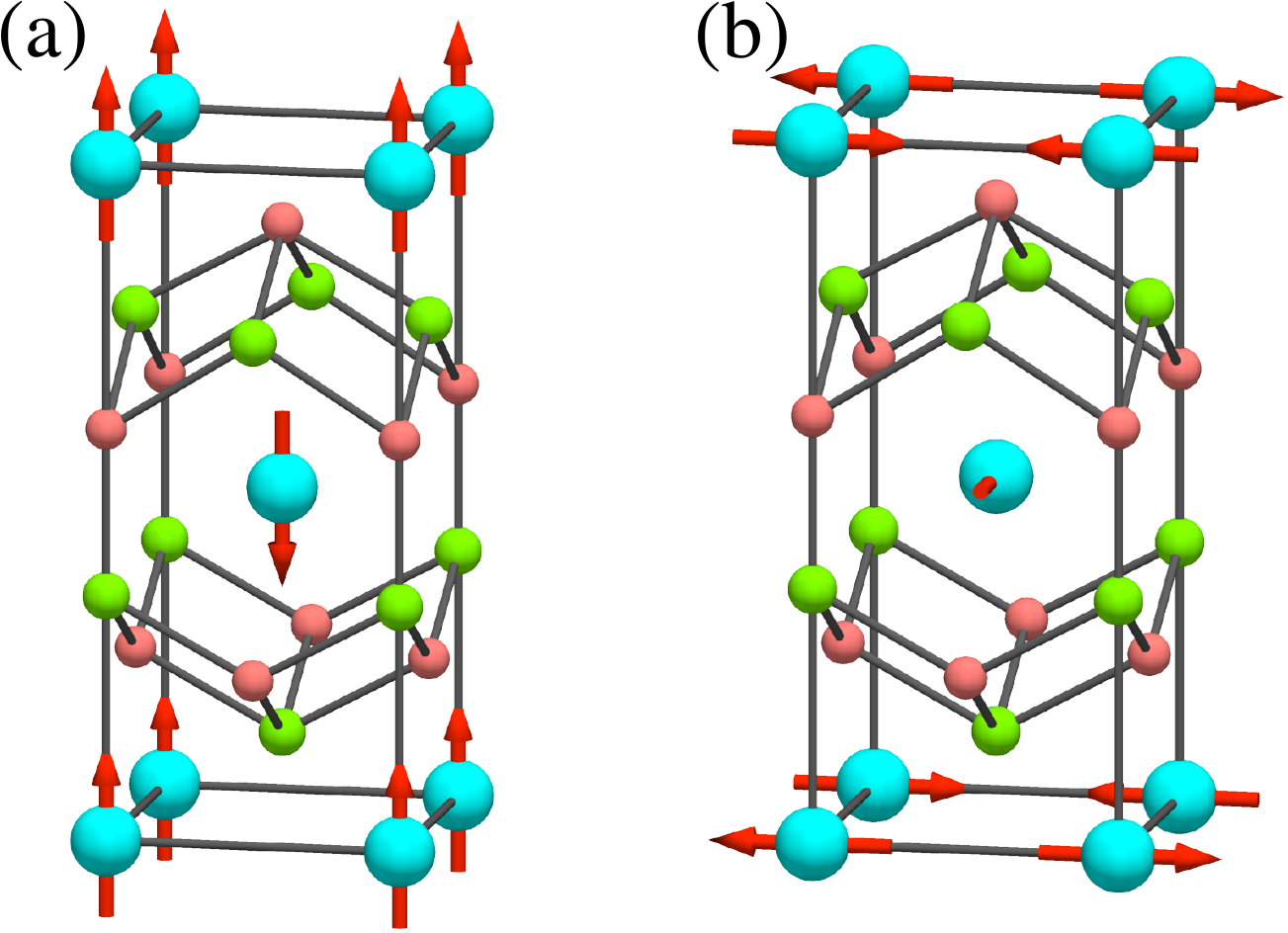}
    \caption{Magnetic structures in the SC1+AFM phase proposed from NQR experiments~\cite{Kibune2022}.
    (a) An AFM structure at $\bm{q}=\QG$, and (b) a helical structure at $\bm{q}=\QA$. The arrows indicate the magnetic dipole.}
    \label{fig:AFM_cand}
\end{figure}

Here, we comment on two magnetic configurations in the SC1+AFM phase proposed from NQR experiments~\cite{Kibune2022}, assuming that the susceptibilities computed in the normal state remain applicable to the superconducting phase.
One is the staggered AFM configuration depicted in Fig.~\ref{fig:AFM_cand}(a), which corresponds in our notation to $M_z(\mathrm{AF})$ with the translation vector $\bm{q}=\QG$.
In our results (Fig.~\ref{fig:chi30-2}), however, the fluctuations at $\bm{q}=\QG$ exhibit no pronounced enhancement, which does not support the AFM configuration in Fig.~\ref{fig:AFM_cand}(a).
The other configuration proposed in the NQR study~\cite{Kibune2022} is a helical structure of in-plane magnetic moments, illustrated in Fig.~\ref{fig:AFM_cand}(b).
This structure can be expressed as a linear combination of $E^-$(F) and $E^-$(AF), namely, $M_x(\mathrm{AF})-M_x(\mathrm{F})-M_y(\mathrm{F})-M_y(\mathrm{AF})$ with $\bm{q}=\QA$.
In Fig.~\ref{fig:chi30-2}, the corresponding fluctuations appear as the second leading instability. Thus, within our DFT+DMFT calculations, the helical magnetic structure shown in Fig.~\ref{fig:AFM_cand}(b) appears as a subleading candidate.

\subsection{Origin of the multipolar fluctuations}

\begin{figure}[t]
    \includegraphics[width=\linewidth]{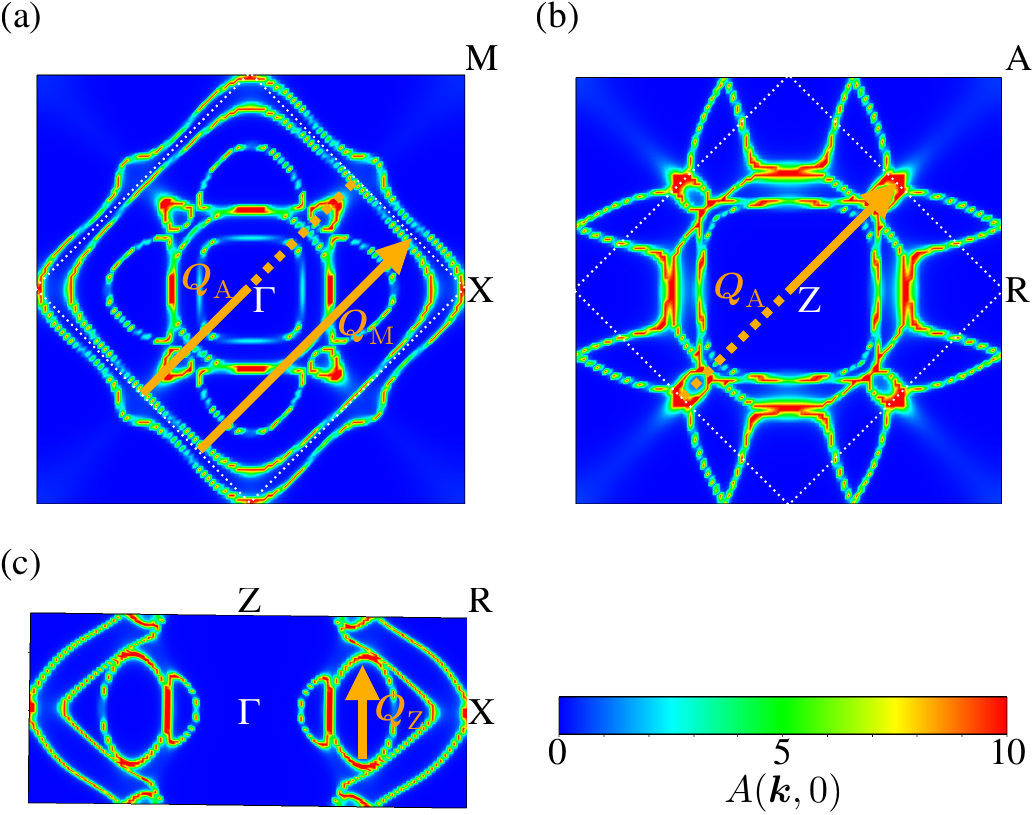}
    \caption{The single-particle excitation spectrum $A(\bm{k},\omega)$ at $\omega=0$ for (a) $k_z=0$, (b) $k_z=\pi/c$, and (c) $k_y=0$. The arrows indicate representative translation vectors, $\QM$, $\QA$, and $\QZ$. Here, $\QA$ connects the Fermi surfaces at $k_z=0$ and $k_z=\pi/c$. The white dashed lines in (a) and (b) indicate half of the Brillouin zone area.}
    \label{fig:fermi}
\end{figure}

Finally, we discuss the origin of the fluctuations. Our SCL formula describes both superexchange and Ruderman-Kittel-Kasuya-Yoshida (RKKY) interactions, depending on  or not the correlated electrons themselves form the Fermi surface, respectively, as demonstrated analytically~\cite{Otsuki2019-SCL}. In the present calculations, the $4f$ electrons are well localized, and the fluctuations obtained above are therefore attributed to the RKKY interaction.
The RKKY interaction, which arises from fluctuations mediated by conduction electrons, is governed by the Fermi surface of conduction electrons.
Figure~\ref{fig:fermi} shows the single-particle excitation spectrum $A(\bm{k},\omega)$ at $\omega=0$ for several cuts in the Brillouin zone. The resulting Fermi surface is consistent with the previous DFT results for {\cerhas}, where the Ce $4f$ electrons are treated as core states~\cite{Chen2024-neutron}. As pointed out in Ref.~\cite{Chen2024-neutron}, this Fermi surface exhibits a nesting property at $\bm{q}=\QM$, as indicated by the arrow in Fig.~\ref{fig:fermi}(a), which accounts for the strong intensities  observed in neutron scattering experiments.
Similarly, the translation vector $\bm{q}=\QM$ connects regions of strong spectral weight between the $k_z=0$ and $k_z=\pi/c$ cuts, as indicated in Figs.~\ref{fig:fermi}(a) and \ref{fig:fermi}(b).
The vector $\bm{q}=\QZ$ corresponds to a connection of the Fermi surface in the $k_y=0$ cut shown in Fig.~\ref{fig:fermi}(c).

The fluctuations at the three translation vectors $\QM$, $\QA$, and $\QZ$ can be understood in terms of the Fermi-surface structure discussed above. In contrast, which multipole components are enhanced can only be determined numerically.

\section{Summary}
\label{sec:Summary}

We derived the momentum-dependent multipolar susceptibilities using the DFT+DMFT method, assuming localized $4f$ electrons.
The energy of the $4f$ level and the interaction parameters were determined from photoemission spectra.
The susceptibilities comprise 72 eigenmodes, classified into 36 atomic multipoles, each associated with ferroic or antiferroic configurations of two Ce sites in the unit cell.
The symmetry of these eigenmodes was identified using a complete set of symmetry-adapted multipole bases.

Among all the fluctuation modes, the susceptibility corresponding to the two-dimensional checkerboard configuration of the $M_z$ magnetic moment is the largest.
This mode is consistent with the inelastic neutron scattering experiments~\cite{Chen2024-neutron}.
Once higher-order multipoles arising from the hybridization between the CEF ground-state and first-excited doublets are included, additional candidates appear as possible order parameters in phase~I if one relies solely on the susceptibility amplitudes. However, most of these candidates can be ruled out by requiring consistency with the characteristic anisotropy of the experimental $T$--$H$ phase diagram.
We thus concluded that the $M_z$ dipole with $\bm{q}=\QM\equiv(1/2,1/2,0)$ is the most plausible, as it exhibits both the largest susceptibility and the correct anisotropic response to the magnetic field.
The increase of $T_0$ under $H \perp c$ is attributed to the enhanced fluctuations of the field-induced quadrupole $\{ Q_{yz}, Q_{zx} \}$, whereas no enhancement is expected for $H \parallel c$ because the field-induced quadrupole $Q_{3z^2-r^2}$ is not locally active.
A subleading candidate is the antiferromagnetic order of the in-plane moment $M_{x+y}$ with $\bm{q}=\QZ\equiv(0,0,1/2)$. 
Orderings of higher-order multipoles, including electric quadrupoles, do not reproduce the observed magnetic-field anisotropy in our DFT+DMFT results.

These results demonstrate an effective use of first-principles approaches in identifying the order parameter of multipolar order in realistic materials.

\begin{acknowledgments}
We thank H. Kusunose for support regarding the MultiPie
library.
This work was supported by JSPS KAKENHI Grants No.~23H04869 and No.~25K22013.
Part of the computations in this work were performed using the facilities of the Supercomputer Center, the Institute for Solid State Physics, the University of Tokyo (Grant No.~2024-Bb-0043 and No.~2025-Ca-0086).
\end{acknowledgments}

\appendix

\section{Symmetry classification of cluster multipoles}
\label{app:multipole}

In the main text, the ferroic and antiferroic configurations within a unit cell are denoted by F and AF, respectively.
When two multipoles are aligned in an AF manner, the symmetry of the resultant cluster multipole is different from the symmetry of the atomic multipole.
Here, a cluster multipole refers to the combined object composed of multipoles within the unit cell.
Classifying the symmetry of such cluster multipoles allows us to predict responses of the ordered state to external fields~\cite{Hayami-review}.

Table~\ref{tab:cluster_multipoles} shows the symmetry classification of the cluster multipoles derived using MultiPie
~\cite{multipie}. The atomic multipoles correspond to those listed in Table~\ref{tab:atomic_multipoles}.
For the F configuration, the symmetry of the cluster multipoles is the same as the atomic multipole, apart from the additional subscript $g$ indicating invariance under spatial inversion.
For the AF configuration, the spatial inversion is odd, indicated by the subscript $u$.

\begin{table*}[!t]
    \caption{Cluster multipoles and their symmetry in {\cerhas}. The ``Atomic'' column lists the atomic multipoles, and ``Irrep(a)'' gives the irreducible representation based on the site symmetry. The ``F/AF" column denotes ferroic (F) and antiferroic (AF) configuration within the unit cell. The last two columns show the symbol of the cluster multipoles and the irreducible representation in the point group $D_{4h}$.}
    \label{tab:cluster_multipoles}
    \begin{tabular}{lllll}\hline
        Irrep(a) & Atomic & F/AF & Cluster & Irrep(c) \\
        \hline
        $A_{1}^+$ & $Q_{0},Q_{3z^2-r^2}$ & F & $Q_{0},Q_{3z^2-r^2}$ & $A_{1g}^+$ \\
        & & AF & $Q_{z}$ & $A_{2u}^+$ \\
        $B_{1}^+$&$Q_{x^2-y^2}$ & F & $Q_{x^2-y^2}$ & $B_{1g}^+$ \\
        & & AF & $G_{xy}$ & $B_{2u}^+$ \\
        $B_{2}^+$&$Q_{xy}$ & F & $Q_{xy}$ & $B_{2g}^+$ \\
        & & AF & $G_{x^2-y^2}$ & $B_{1u}^+$ \\
        $E^+$&$\{ Q_{yz},\ Q_{zx} \}$ & F & $\{ Q_{yz},\ Q_{zx} \}$ & $E_{g}^+$\\
        & & AF & $\{ Q_{x},\ Q_{y} \}$ & $E_{u}^+$ \\
        \hline
        $A_{2}^-$&$M_{z},M_{z(5z^2-3r^2)}$ & F & $M_{z},M_{z(5z^2-3r^2)}$ & $A_{2g}^-$ \\
        & & AF & $M_{0},M_{3z^2-r^2}$ & $A_{1u}^-$ \\
        $B_{1}^-$&$M_{xyz}$ & F & $M_{xyz}$ & $B_{1g}^-$ \\
        & & AF & $M_{xy}$ & $B_{2u}^-$ \\
        $B_{2}^-$&$M_{z(x^2-y^2)}$ & F & $M_{z(x^2-y^2)}$ & $B_{2g}^-$ \\
        & & AF & $M_{x^2-y^2}$ & $B_{1u}^-$ \\
        $E^-$&$\{ M_{x},\ M_{y} \},$ & F & $\{ M_{x},\ M_{y} \},$ & $E_{g}^-$ \\
        &$\{ M_{x(x^2-z^2)},\ M_{y(y^2-z^2)} \},$ &  & $\{ M_{x(x^2-z^2)},\ M_{y(y^2-z^2)} \},$ & \\
        &$\{ M_{x(5x^2-3r^2)},\ M_{y(5y^2-3r^2)} \}$ &  & $\{ M_{x(5x^2-3r^2)},\ M_{y(5y^2-3r^2)} \}$ & \\
        && AF & $\{ T_{x},\ T_{y} \}$ & $E_u^-$ \\
        & &  & $\{ M_{yz},\ M_{zx} \}$ &  \\
        &&  & $\{ T_{x(5x^2-3r^2)},\ T_{y(5y^2-3r^2)} \}$ &  \\
        \hline
    \end{tabular}
\end{table*}

\begin{table*}[!t]
      \centering
      \caption{Candidates of the quadrupole or octupole order parameter $\phi_1$ in phase~I, and the field-induced multipoles $\phi_2$. The enhancement factor $\alpha_1$ is defined in Eq.~\eqref{eq:enhance1}.
      The dash in the column for $\alpha_1$ means that Eq.~\eqref{eq:enhance1} is not applicable because of $I_2(\bm{q})/I_1(\bm{q})>1$.
      See also the caption of Table~\ref{tab:enhance_dipole}.}
      \label{tab:enhance_quadrupole}
      \begin{tabular}{ccc|cccc|cccc}\hline
            $\phi_1$         & F/AF & $\bm{q}$ & Field direction & $\phi_2$           & $I_2/I_1$ & $\alpha_1$ & Field direction & $\phi_2$           & $I_2/I_1$ & $\alpha_1$ \\\hline
            $M_{z(x^2-y^2)}$ & AF   & $\QZ$    & [001]           & $Q_{x^2-y^2}$      & $0.85$    & $6.5$      & [100]           & $Q_{zx}$           & $-0.17$   & $0.86$     \\\hline
            $Q_{x^2-y^2}$    & AF   & $\QZ$    & [001]           & $M_{z(x^2-y^2)}$   & $1.2$     & --         & [100]           & $M_{x(y^2-z^2)}$   & $0.15$    & $1.2$      \\
                             &      &          &                 &                    &           &            &                 & $M_{x(5x^2-3r^2)}$ & $0.63$    & $2.7$      \\\hline
            $Q_{zx}$         & F/AF   & $\QA$    & [001]           & $M_{x(y^2-z^2)}$   & $0.40$    & $1.7$      & [100]           & $M_{z(x^2-y^2)}$   & $-0.16$   & $0.86$     \\
                             &      &          &                 & $M_{x(5x^2-3r^2)}$ & $0.62$    & $2.6$      &                 & $M_{z(5z^2-3r^2)}$ & $-0.14$   & $0.87$     \\\hline
      \end{tabular}
\end{table*}

\section{Anisotropy of quadrupolar ordering in pseudo-quartet model}
\label{app:anisotropy}

Section~\ref{sec:anisotropy} discusses the anisotropy of the magnetic-field enhancement of the transition temperatures for the magnetic dipole order. Here, we apply the same analysis to the electric quadrupoles and magnetic octupoles, which arise from hybridization between the CEF ground-state and first-excited doublets.

Table~\ref{tab:enhance_quadrupole} shows the quadrupole and octupole counterparts of Table~\ref{tab:enhance_dipole}.
The three multipoles exhibiting the largest fluctuations in Fig.~\ref{fig:chiir_2nd} are included.
When $\phi_1$ corresponds to the electric quadrupole, the induced moment $\phi_2$ is the magnetic octupole, and vice versa.
The ratio $I_2/I_1$ is obtained from the results shown in Fig.~\ref{fig:Iq}.
The enhancement factor $\alpha_1$ is evaluated using Eq.~\eqref{eq:enhance1}, with the assumption that the two CEF doublets are degenerate and give rise to the Curie-Weiss behavior of the susceptibilities.
We find, for all three cases, that the ratio $I_2/I_1$ is smaller for $H \perp c$ than for $H \parallel c$, resulting in a smaller enhancement factor for $H \perp c$.
This tendency is opposite to the experiments, where the transition temperature $T_0$ is enhanced (suppressed) for $H \parallel c$ ($H \perp c$).
Therefore, the DFT+DMFT results do not support the electric-quadrupole and magnetic-octupole scenario in the pseudoquartet CEF system for phase~I.

\bibliography{refs}

\end{document}